\documentclass[preprint]{aastex}

\newcommand{\citar}[1]{\citeauthor{#1} (\citeyear{#1})}
\newcommand{\citarwiths}[1]{\citeauthor{#1}'s (\citeyear{#1})}

\newcommand{\citartwo}[2]{\citeauthor{#1} (\citeyear{#1}, \citeyear{#2})}

\newcommand{\citarNP}[1]{\citeauthor{#1} \citeyear{#1}}
\newcommand{\citarNPwiths}[1]{\citeauthor{#1}'s \citeyear{#1}}

\newcommand{\citartwoNP}[2]{\citeauthor{#1} \citeyear{#1}, \citeyear{#2}}

\newcommand{\citarfourNP}[4]{\citeauthor{#1} \citeyear{#1}, \citeyear{#2},
\citeyear{#3}, \citeyear{#4}}

\newcommand{\threej}[6]
{ \left( \begin{array}{ccc}
#1&#2&#3\\
#4&#5&#6
\end{array} \right) }

\renewcommand{\case}[4]
{ \Bigg\{
\begin{array}{cc}
#1 & #2 \\
#3 & #4
\end{array} }

\newcommand{\casesmall}[4]
{ \Big\{
\begin{array}{cc}
#1 & #2 \\
#3 & #4
\end{array} }

\newcommand{\rotmat}[3]{\mathcal{D}^{#1}_{#2 #3}}

\begin{document}
\title{Theory and Modeling of the Zeeman and Paschen-Back effects in Molecular Lines}

\author{A. Asensio Ramos, J. Trujillo Bueno\altaffilmark{1}}
\affil{Instituto de Astrof\'{\i}sica de Canarias, 38200, La Laguna,
Spain}
\altaffiltext{1}{Consejo Superior de Investigaciones Cient\'{\i}ficas, Spain}
\email{aasensio@iac.es, jtb@iac.es}

\begin{abstract}

This paper describes a very general approach to the calculation of the 
Zeeman splitting effect produced by an external magnetic field on the rotational
levels of diatomic molecules. The method is valid for arbitrary values of the
total electronic spin and of the magnetic field strength -that is, it holds for
molecular electronic states of any multiplicity and for both the Zeeman and
incomplete Paschen-Back regimes. It is based on an efficient numerical
diagonalization of the effective Zeeman Hamiltonian, which can incorporate
easily all the contributions one may eventually be interested in, such as the
hyperfine interaction of the external magnetic field with the spin motions of
the nuclei. The reliability of the method is demonstrated by
comparing our results with previous ones obtained via formulae valid only for
doublet states. We also present results for molecular transitions arising
between non-doublet electronic states, illustrating that their Zeeman patterns
show signatures produced by the Paschen-Back effect.

\end{abstract}


\section{Introduction}

Most polarized radiation diagnostics 
of astrophysical magnetic fields have been carried out via the
theoretical interpretation of the observed polarization signatures in {\em
atomic} spectral lines (e.g., the reviews by \citarNP{bagnulo_spw3_03}; 
\citarNP{mathys02}; \citarNP{stenflo02}). 
However, over the last few years we have witnessed an increasing
interest in molecular spectropolarimetry as a tool for empirical
investigations on solar and stellar magnetism, 
concerning both the molecular Zeeman effect (e.g., the recent overviews by \citarNP{asensio_trujillo_spw3_03}; 
\citarNP{berdyugina_spw3_03} and \citarNP{landi_AN_03}; see also \citarNP{uitenbroek_asensio04}) and the Hanle effect in 
molecular lines (e.g., \citarNP{egidio03}; \citarNP{trujillo_spw3_03}).

The aim of the present paper is to describe in some detail our
approach to the Zeeman and Paschen-Back effects
in diatomic molecular lines, which we have been applying
over the last few years while contributing to the development
of the field of radiative transfer in molecular lines
(\citartwoNP{asensio_trujillo_spw3_03}{asensio_trujillo05c};  
\citarfourNP{asensio03}{asensio_ch04}{asensio_trujillo04}{asensio_trujillo05a}). 
Our strategy is based on a a very efficient numerical diagonalization of the
effective Zeeman Hamiltonian, whose general expression has been derived by a
number of authors
working in the field of molecular physics (e.g., \citarNP{brown_carrington03}).
Interestingly,
in physics laboratory experiments, where the applied magnetic
field is known beforehand, the observation of the splittings provide a
measurement of the magnetic dipole moment of the considered molecule in the
particular spin-rotational level involved. Given that
the magnetic moment depends on the electronic structure of the
molecule, it is obvious that its measurement provides information about the
molecular structure. In astrophysics we have the inverse problem, the magnetic
field being the unknown quantity. To obtain information about cosmic magnetic
fields, therefore, we have to make use of our most precise knowledge on
molecular structure in order to infer the magnetic field vector via the
theoretical modeling of polarization signals in molecular lines. To this end, we
need:

\begin{itemize}
\item to obtain the molecular number densities at each point within
the astrophysical plasma model under consideration, 

\item to calculate the
splittings of the molecular energy levels with the strengths of the
individual Zeeman components, and 

\item to solve the radiative transfer
problem for the emergent Stokes parameters. 

\end{itemize}

Here we focus on the issue of
calculating the splittings and strengths of molecular transitions, including the
most general situation of the incomplete Paschen-Back effect for electronic
states of arbitrary multiplicity. For information on the techniques we use for
calculating the molecular number densities and for solving the Stokes-vector
radiative transfer problem in (magnetized) stellar atmospheres we refer the
reader to the papers by \citar{asensio03} and \citar{trujillo03},
respectively.

It is of historical interest to mention that
the Zeeman effect in diatomic molecules was considered shortly after the
development of the quantum theory. \citar{kronig28}
investigated the molecular Zeeman effect in 
Hund's (a) and (b) cases for
the angular momentum coupling between electronic and rotational motion.
Only one year after, \citar{hill29} investigated the Zeeman effect for doublet
states 
of diatomic molecules in intermediate states between Hund's cases (a) and (b).
The status of the theory was reviewed by \citar{crawford34},
emphasizing that, at that moment,
the Zeeman effect was understood in pure Hund's cases (a) and (b) and in
intermediate
cases between the two. At that time, the intensity of the
Zeeman transitions had not been
investigated in detail because the calculations were rather involved when using
the
basis functions of Hund's case (b). Fifty years after the paper of Kronig, \citar{schadee78}
re-investigated the Zeeman effect
for doublet states of diatomic lines in the intermediate case between
Hund's cases (a) and (b), but using the basis functions of Hund's case (a). This greatly 
facilitated the calculation of 
the intensity of the Zeeman transitions. The issue of the Zeeman effect in lines
of diatomic molecules was later considered by
\citar{illing81} who studied in great detail Schadee's (1978) theory
and applied it for understanding the broad-band \emph{circular} polarization
observed by \citar{harvey73} in near-IR lines of CN.

All the above-mentioned developments were based on a number of 
approximations for the description of the molecular motions for the zero
field case. They
take into account the rotational energy (but neglecting centrifugal distortions)
and the strongest angular momenta couplings.
For states without electronic orbital angular momentum
($\Sigma$ states), only the spin-rotation coupling was included. For states with
non-zero electronic orbital angular momentum 
($\Pi$, $\Delta$, \ldots), only the spin-orbit coupling was taken into account.

The formulae developed by \citar{schadee78} are only applicable
to doublet states of diatomic molecules in the Zeeman or Paschen-Back regimes.
Recently, \citar{berdyugina02a} have extended \citarwiths{schadee78} formulation
to allow for the calculation of the effect of a magnetic field on states with
arbitrary spin, but limited to the Zeeman regime. 
Their strategy consisted in numerically diagonalizing a simplified Hamiltonian
corresponding to the
zero-field case and accounting for the Zeeman Hamiltonian as a first order
perturbation. This approach, which neglects the
non-diagonal matrix elements ($\Delta J \neq 0$) of the Zeeman Hamiltonian, is
only
valid in the linear Zeeman regime. Therefore, 
the maximum reliable value of the magnetic field strength 
is established by the transition to the Paschen-Back regime.

In this paper we present a very
general approach which allows us 
to calculate the effect of a magnetic field on the
rotational levels of diatomic molecules in electronic states with arbitrary
multiplicity.
The method is valid in both the Zeeman and the
Paschen-Back regimes. It is based on the numerical diagonalization of the
effective molecular Hamiltonian, which
describes the molecular motion using the basis functions of Hund's case (a).
Therefore, the inclusion of any
additional contribution to the effective Hamiltonian of the diatomic molecule is
straightforward, provided the matrix elements of the Hamiltonian are known in
the basis functions of Hund's
case (a). This makes it possible to investigate effects like hyperfine structure
(HFS) without much additional effort. Such refinements
are not only of interest in physics laboratory experiments, but also when it
comes to interpreting correctly the linear polarization signals that anisotropic
radiation pumping processes induce in spectral lines (e.g., \citarNP{landi_landolfi04}). 
The illustrative examples shown in this paper neglect HFS
effects, but we plan to
consider this interesting molecular HFS problem in a future investigation.

The outline of this paper is the following.
Section 2 and the various appendices describe our numerical diagonalization
approach in some detail, including a summary
of some results from angular momentum theory with the aim of 
facilitating a better understanding of the paper. 
In order to verify the reliability
of our numerical results, in Section 3 we compare 
them with those that can be obtained via Schadee's (1978) theory. We will show
also some applications to non-doublet
states. Finally, Section 4 summarizes our main conclusions with an outlook to
future research.

\section{Angular momentum theory for diatomic molecules}
The theory of angular momentum provides a robust theoretical framework for
describing the complex structure
of diatomic molecules. In this section we summarize how the 
angular momentum theory and Racah's tensor
algebra allow the description of the quantum mechanical behavior of any
diatomic molecule under
the presence of an arbitrary magnetic field. Although the theory on which our
work is based on can be found explained in detail in
specialized monographs on molecular spectroscopy (e.g., \citarNP{judd75},
\citarNP{brown_carrington03}), we have decided to present here a
brief summary in order to facilitate the 
understanding of this paper and to avoid any possible notational confusion.

\subsection{Introduction}
As is well-known,
the Born-Oppenheimer approximation leads to an effective separation of the
energies due to the electronic, vibrational
and rotational motions in molecules (e.g., Herzberg 1950). 
This approximation is supported by the fact that the mass of the nuclei is
several orders
of magnitude larger than the mass of the electrons. This implies that, when
nuclei move, the
electrons adapt rapidly to the new nuclear configuration. As a result, the
electrons feel some kind of effective potential,
which depends only on the positions of the nuclei and on the electronic
configuration
of the particular electronic state. The Born-Oppenheimer approximation
inmediately leads to the possibility
of separating the eigenfunctions associated with each motion. 

For convenience, we have selected
Hund's case (a) eigenfunctions because they lead to several simplified
expressions (see Appendix \ref{app_hund_cases}
for details). In particular, the eigenvalues of the angular momentum operators
are very easy to obtain in
this basis set, which greatly simplifies the calculation of the matrix elements
of the
total Hamiltonian. The obvious consequence of selecting a 
basis set, which is a truly good basis only in 
some limiting conditions, is that the total Hamiltonian
is usually non-diagonal and it has to be diagonalized to obtain the
energies of the molecular levels.
In the usual notation, the Hund's case (a) eigenfunctions can be written as
(\citarNP{judd75}):
\begin{equation}\label{eq_5_1}
|\alpha \Lambda S \Sigma ; v ; \Omega J M \rangle \equiv |\alpha \Lambda S
\Sigma
\rangle | v \rangle | \Omega J M \rangle,
\end{equation}
where we have explicitly indicated that the total eigenfunction is a product of
the eigenfunctions
associated with the electronic, vibrational and rotational motions (which is
only strictly valid under
the Born-Oppenheimer approximation). 

We point out that $|\alpha \Lambda S \Sigma \rangle$ is the
electronic eigenfunction. The symbol $\alpha$ represents a collection
of quantum numbers which are used to label the electronic configuration of the
molecule, while $S$ is the total spin.
The symbols $\Lambda$ and $\Sigma$ are the projections of the total orbital
electronic angular momentum ($\mathbf{L}$) and of the
total spin ($\mathbf{S}$) on the internuclear axis, respectively. Both are good quantum
numbers in Hund's case (a), while $\Sigma$ is not
a good quantum number in Hund's case (b) (see Appendix \ref{app_hund_cases} for
more details).
The vibrational eigenfunction is represented by $|v \rangle$. Because we are
interested in lines of diatomic molecules, this eigenfunction
can be correctly described with only one quantum number $v$. Finally,
$| \Omega J M \rangle$ represents the rotational eigenfunction. This function
depends on the
total angular momentum $J$, its projection $M$ on the quantization axis
(typically chosen as the axis along the 
magnetic field vector) and the number $\Omega=|\Lambda|+\Sigma$. It is
interesting to note that
$\Omega$ is a good quantum number only in Hund's (a) case, since
$\Sigma$ is only well defined in this
case. Note also that, due to the presence of the absolute value, in
the non-rotating molecule the energy of the 
spin-orbit levels with $\Omega \geq 1/2$ are doubly degenerate. When the molecule is rotating, this
degeneracy is broken (see, e.g. \citarNP{herzberg50}).
A brief summary of the good quantum numbers in Hund's cases (a) and (b) can be
found in 
Table \ref{tab_good_numbers}. Additionally, vector diagrams of the two coupling
cases are shown in 
Figure \ref{fig_hunds_a}, which clearly show the strength of each coupling.

\begin{figure}
\plotone{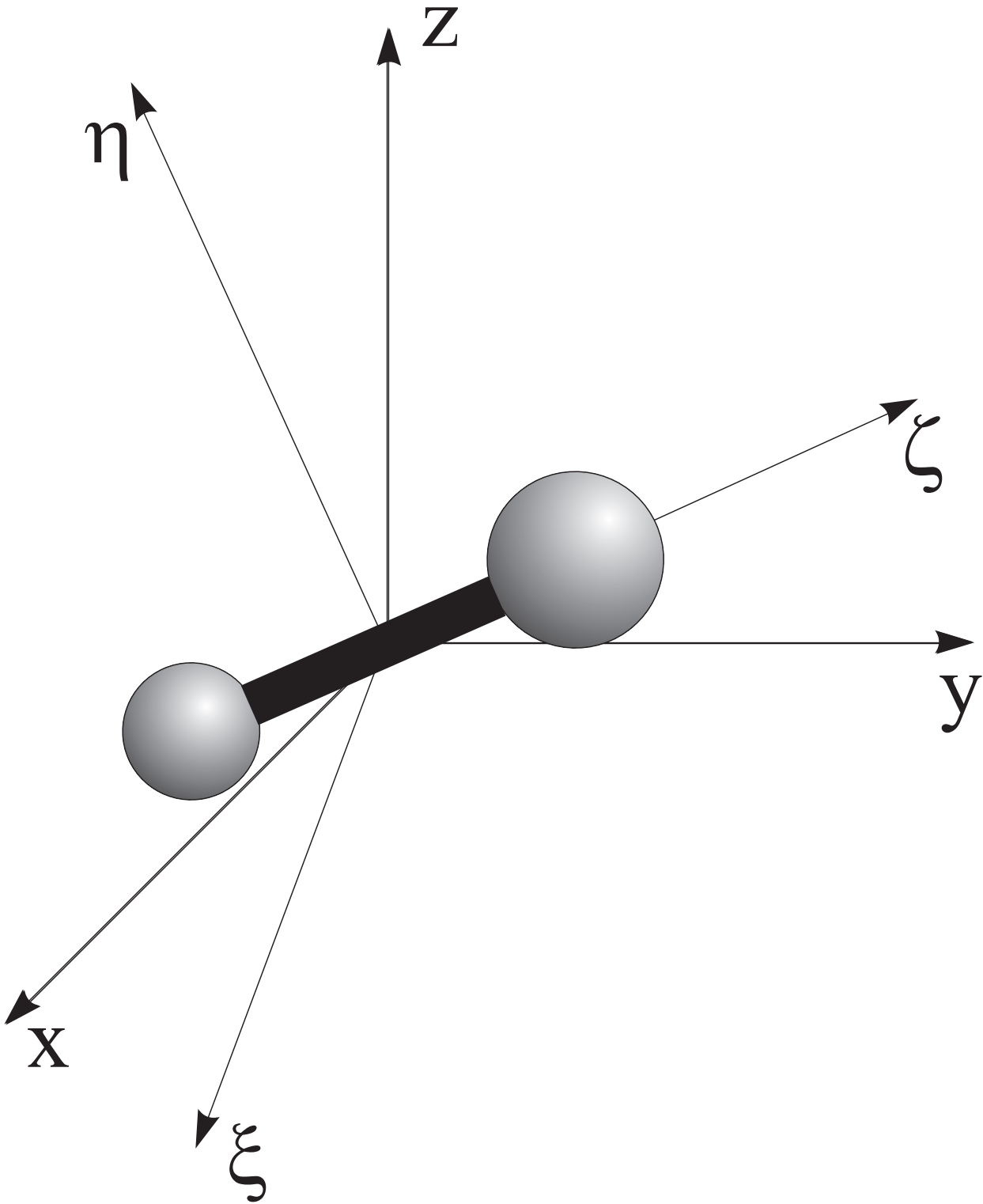}
\caption{The two frames used to describe the motion of a diatomic molecule. The
$F$ frame with axes $(x,y,z)$ is fixed at the center of mass of the
molecule, with the $z$ axis along the quantization axis. The $F'$ frame with
axes $(\xi,\eta,\zeta)$
rotates with the molecule, with the $\zeta$ axis
along the internuclear axis. The transformation of any tensorial quantity
between both frames can be easily carried out with the aid of
Eqs. (\ref{eq_5_6}) and (\ref{eq_5_6b}).}
\label{fig_geometry}
\end{figure}

Due to the difference in mass between the electrons and the nuclei of the
diatomic
molecule, it is appropriate to refer the motion of
the electrons to a frame $F'$ fixed to the nuclei whose origin is at the center of
mass of the molecule, rather than to an external
laboratory frame $F$ (with its origin also at the center of mass of the
molecule). Fig. \ref{fig_geometry} illustrates the relative position of both frames. The
coordinates of an electron in the laboratory frame $F$ are $(x,y,z)$, 
while the coordinates on the frame fixed to the
molecule $F'$ are $(\xi,\eta,\zeta)$. Given the arbitrariness in the choice
of the frames, it is convenient to consider the $\zeta$ axis of the $F'$ frame
parallel to the internuclear axis and the $z$ axis of the $F$ frame along the quantization axis.
This
quantization axis is usually chosen along the direction of the magnetic field
vector when it is present. Obviously, in the absence
of a magnetic field, it can be arbitrarily chosen in any direction.
Since both reference frames share the same origin, coordinates in one frame can be transformed to the other
frame by means of
a simple rotation. This rotation can be parameterized in terms of the three
Euler
angles $(\alpha,\beta,\gamma)$. Once the coordinates of an electron are known in
one of
the frames,
a standard geometrical rotation between both frames using the Euler angles can
be used to
obtain the coordinates in the other frame.
As usual, once the $\zeta$ axis of the $F'$ frame is fixed, there is an
additional freedom in the choice of
the direction of the $\xi$ and $\eta$ axis, which
depends on our choice for the Euler angle $\gamma$. Without loss of generality, we
choose $\gamma=\pi/2$.

The normalized rotational eigenfunctions $| \Omega J M \rangle$ in Hund's case
(a) can be obtained by solving
the Schr\"odinger equation for a symmetric top. The solution depends explicitly
on the Euler angles, and
can be written as (\citarNP{judd75})
\begin{equation}\label{eq_5_2}
| \Omega J M \rangle = \sqrt{\frac{2J+1}{8 \pi^2}} \rotmat{J}{M}{\Omega}
(\alpha,\beta,\gamma)^*,
\end{equation}
where $\rotmat{J}{M}{\Omega}(\alpha,\beta,\gamma)$ is the rotation matrix
(e.g., \citarNP{edmonds60}). The
representation of the rotational eigenfunctions in terms of the rotation
matrices is very convenient because it facilitates the subsequent calculations.
A fundamental property is that of conjugation 
for the rotation matrices, which can be expressed as
\begin{equation}\label{eq_5_3}
\rotmat{J}{M}{\Omega} (\alpha,\beta,\gamma)^* = (-1)^{M-\Omega}
\rotmat{J}{-M}{-\Omega} (\alpha,\beta,\gamma).
\end{equation}
Other important properties of the rotation matrices which are used in this
section are summarized in Appendix B. Note that the conjugation property allows
us to verify that the rotational eigenfunctions are orthonormal. In fact,
one can apply Eq. (\ref{eq_5_3}) together with Eq. (\ref{eq_5_5}) to verify
that 
$\langle \Omega' J' M' | \Omega J M \rangle = \delta_{\Omega' \Omega} \delta_{J'
J} \delta_{M' M}$.

The rotation matrices are of great help for transforming tensorial operators
from one
frame to another, provided that the Euler angles between
both frames are known. Given that we are dealing with rotations, it is
advantageous to consider the spherical components
of a tensor instead of working with the cartesian ones (see, e.g.,
\citarNP{edmonds60} for the
definition of the spherical components of a tensor). Let us consider in some
detail the case of a vector, since it will be helpful for calculating the matrix
elements of the Hamiltonian.
Assume that $\mathbf{r}$ is a 
vector in $\mathbb{R}^3$ whose spherical components in the laboratory frame $F$
are given 
by $\bar{r}_q$, where $q=0,\pm 1$. The spherical components of the same vector
in the
frame fixed to the molecule $F'$ are denoted by $r_q$. The relationship between
both components can be written as (e.g., \citarNP{edmonds60}):
\begin{equation}\label{eq_5_6}
\bar{r}_q = \sum_{q'=0,\pm 1} r_q \rotmat{1}{q'}{q} (-\alpha,-\beta,-\gamma),
\end{equation}
where we have used the fact that the rotation needed to carry the frame
fixed to
the molecule to the laboratory frame is
expressed by the inverse rotation $(-\alpha,-\beta,-\gamma)$. Using the
properties
of the rotation matrices (see, e.g., \citarNP{edmonds60}), the previous equation
can also be written as:
\begin{equation}\label{eq_5_6b}
\bar{r}_q = \sum_{q'=0,\pm 1} r_q \rotmat{1}{q}{q'} (\alpha,\beta,\gamma)^*.
\end{equation}

The final part of this introduction to the mathematical 
tools needed to calculate the matrix elements
of the Hamiltonian is related to the Wigner-Eckart (WE) theorem (e.g.,
\citarNP{edmonds60}). This theorem facilitates the calculation
of the matrix element of any tensor operator because it takes full advantage
of any symmetry that may be inherent to the problem under consideration.
In other words, it isolates those parts of a problem that are essentially
geometric in character from those which depend
explicitly on the physics of the problem. For the $q$ component of a given
tensor of rank $k$, the WE theorem reads:
\begin{equation}\label{eq_5_7}
\langle J M | T_q^{(k)} | J' M' \rangle = (-1)^{J-M}
\threej{J}{k}{J'}{-M}{q}{M'} \langle J \| T^{(k)} \| J' \rangle.
\end{equation}
The quantity $\langle J \| T^{(k)} \| J' \rangle$ is called the reduced matrix
element,
a number which depends on the physics of the
selected problem. An example is the reduced matrix element that involves an
angular momentum $\mathbf{J}$ between its own
eigenfunctions:
\begin{equation}\label{eq_5_8}
\langle J \| \mathbf{J} \| J \rangle = \sqrt{J(J+1)(2J+1)}
\end{equation}

\subsection{Molecular Hamiltonian}
\label{sec_mol_hamiltonian}
In this section we present the total effective Hamiltonian which we use to
describe the molecular motion including the effect of a magnetic field.
Consider a diatomic molecule in a given electronic, vibrational and rotational
state and in the presence of an external magnetic field.
It is possible to describe the molecular motion by writing the full Hamiltonian
which takes into account all the Coulomb forces among electrons and
nuclei and the Lorentz force due to the presence of a magnetic field. However, it is more 
appropriate in terms of both economy and feasibility to consider the effective
Hamiltonian approach (\citarNP{brown79}, \citarNP{brown_carrington03}). This effective Hamiltonian
has the same eigenvalues as the full Hamiltonian, but it operates only within
the rotational, spin and
hyperfine levels of a given vibrational level belonging to a given electronic
state. The non-diagonal matrix elements (coupling the vibrational 
level of interest with any other vibrational level of any other electronic
state) are absorbed in the effective Hamiltonian and 
represented by some parameters which can be obtained by comparison with
laboratory or astrophysical observations.
Following \citar{brown_carrington03} (see also \citarNP{brown79}) we write down
the
following very general expression for the \emph{effective Hamiltonian}:
\begin{eqnarray}\label{eq_5_21}
H_\mathrm{eff} &=& H_\mathrm{SO} + H_\mathrm{SS} + H_\mathrm{rot} + H_\mathrm{cd}
+ H_\mathrm{sr} + H_\mathrm{LD} \nonumber \\
&+& H_\mathrm{cdLD} + H_\mathrm{hfs} + H_\mathrm{cdhfs} + H_Z.
\end{eqnarray}
The term $H_\mathrm{SO}$ represents the coupling between the total spin of the molecule
$\mathbf{S}$ and the orbital angular
momentum $\mathbf{L}$. $H_\mathrm{SS}$ represents the coupling between the spins
of the electrons in the molecule. The rotational energy of the molecule is accounted for by the term 
$H_\mathrm{rot}$, while
$H_\mathrm{cd}$
represents the contribution to the rotational energy of the centrifugal
distortion as a consequence of the rotation of the
molecule. The interaction between the total molecular spin and the rotational
angular momentum is described by $H_\mathrm{sr}$.
The additional degeneracy present for 
the two possible projections of $\mathbf{L}$ on the internuclear axis, namely $\pm |\Lambda|$, is broken by 
the inclusion of the terms $H_\mathrm{LD}$ and
$H_\mathrm{cdLD}$ of the effective Hamiltonian. These terms represent the
$\Lambda$-doubling interactions (interaction with higher energy electronic states) and the
appropriate centrifugal distortion corrections, respectively.
In case one is interested in taking into account the hyperfine structure, it 
suffices to include the terms $H_\mathrm{hfs}$ and $H_\mathrm{cdhfs}$, which
are the hyperfine interaction and
the corresponding centrifugal distortion of the hyperfine
interaction as the molecule rotates, respectively. Such hyperfine contributions
will not be considered in this paper, but they will be the subject of a future
investigation. Finally, $H_Z$ represents the interaction between the molecule and an external magnetic field $\mathbf{B}$.

Each one of the terms of the Hamiltonian can be written with the aid of the
different angular momentum operators that can be defined in the molecule.
As previously mentioned, it is useful to write all these operators using the
spherical
tensor notation because it leads to considerable simplifications in the notation
and
it allows the use of the powerful tools of the angular momentum theory. The same
notation we have used for vectors can be straightforwardly extended to
the spherical components of tensorial operators. On the one hand, $A^k_q$ will
refer to 
the $q$ spherical component of the tensor
$\mathbf{A}^k$ of rank $k$ in the molecular fixed frame $F'$. On the other
hand, the spherical component
of the same tensor in the laboratory frame $F$ will be indicated by
$\bar{A}^k_q$. This extension to tensors of arbitrary rank $k$ is only
needed when we include in the
Hamiltonian tensorial operators (see, e.g., \citarNP{brown_carrington03}). In
the effective Hamiltonian 
used in this paper we consider only vectorial operators with rank up to $k=1$.
However, it is important to emphasize that this formalism allows us to include
any contribution to the effective Hamiltonian  written in terms of tensorial
operators, which
constitutes one of its big advantages.

The explicit form of each of the terms included in the effective Hamiltonian
has been obtained previously by molecular physicists (see
\citarNP{brown_carrington03}, and references therein). The effective Hamiltonian
approach is flexible enough to allow writing the explicit form of some of the terms of the Hamiltonian
in different ways. It is advantageous to write them in a way that simplifies
the calculation of the matrix elements in the chosen basis set. 
The spin orbit coupling term reads
\begin{equation}\label{eq_5_22}
H_\mathrm{SO} = A L^1_{0} S^1_0 + \frac{1}{2} A_D \left[ \mathbf{N}^2 L^1_0
S^1_0 +
L^1_0 S^1_0 \mathbf{N}^2 \right],
\end{equation}
where $A$ and $A_D$ are the spin-orbit coupling constant and the
centrifugal correction to $A$, respectively. The first part includes the
coupling between the total
spin and the orbital angular momentum. Note that this contribution using
Hund's case 
(a) basis set is just proportional to
the product of $\Lambda$ and $\Sigma$ since these are the $L^1_0$ and $S^1_0$
components of $\mathbf{L}$ and
$\mathbf{S}$, respectively. The second term includes the effect of the
centrifugal distortion in the spin-orbit
coupling. The symbol $\mathbf{N}^2$ stands for the scalar product of
$\mathbf{N}$ by itself, i.e.,
$\mathbf{N}^2=\mathbf{N} \cdot \mathbf{N}$. 
Following \citar{brown_carrington03}, we use $\mathbf{N}^2$ as the square of the rotational operator instead 
of $\mathbf{R}^2$ (see Appendix \ref{app_hund_cases}). The latter approach is preferred by some authors because this 
operator is proportional to the 
rotational kinetic energy. However, using $\mathbf{R}$ as the rotational operator involves the angular momentum $\mathbf{L}$ 
which has non-diagonal elements between different electronic levels. In the effective Hamiltonian approach, the effect of 
these non-diagonal 
terms are absorbed into the rotational and coupling constants. This way, we end up with a Hamiltonian which operates only
within the (rotational, spin and hyperfine) levels of a given vibrational level of an individual electronic state.

Focusing now on the spin-spin interaction, several 
functional forms have been suggested in the literature (e.g., \citarNP{brown_carrington03} for a summary). It has been
shown that they are equivalent
and we have decided to choose the following functional form which greatly simplifies the
calculation of the matrix elements
in Hund's case (a) basis set:
\begin{equation}\label{eq_5_22b}
H_\mathrm{SS} = \frac{2}{3} \lambda \left[ 3 (S^1_0)^2 - \mathbf{S}^2 \right],
\end{equation}
where $\lambda$ is the spin-spin coupling constant. Obviously, this contribution turns out to be
important only when more than one
electron is uncoupled ($S > 1/2$).

Turning our attention to the rotational part of the effective Hamiltonian, the
pure rotation Hamiltonian 
can be written as
\begin{equation}\label{eq_5_23}
H_\mathrm{rot} = B \mathbf{N}^2,
\end{equation}
where $B$ is the usual rotational constant 
For the evaluation of the matrix
elements in the basis set of Hund's
case (a), it is better to represent the rotational angular momentum $\mathbf{N}$
in terms of $\mathbf{J}$ and $\mathbf{S}$ since these operators have well-defined
eigenvalues in this basis set. This transformation can be easily carried out 
because $\mathbf{N}=\mathbf{J}-\mathbf{S}$ (see Appendix \ref{app_hund_cases}). 
Concerning the inclusion of centrifugal distortion terms, it is 
sufficient to arrive up to the third
order contribution even for laboratory 
experiments, and the explicit form of the Hamiltonian turns out to be:
\begin{equation}\label{eq_5_24}
H_\mathrm{cd} = -D (\mathbf{N}^2)^2 + H (\mathbf{N}^2)^3,
\end{equation}
where $D$ and $H$ are the quartic and sextic distortion constants. Finally, the
coupling of the spin and the rotation can be included in the effective
Hamiltonian as the scalar product
between the rotational angular momentum and the spin:
\begin{equation}\label{eq_5_25}
H_\mathrm{sr} = \gamma (\mathbf{J}-\mathbf{S}) \cdot \mathbf{S},
\end{equation}
where $\gamma$ is the spin-rotation coupling constant (note that this constant
is different from the $\gamma$ Euler angle). In this paper
we will not include more terms in the effective Hamiltonian for the zero-field case.

An additional constraint has to be
included when dealing with diatomic molecules that greatly simplifies the evaluation of the matrix elements of the
effective Hamiltonian.
This constraint is related to the fact that the molecule does not rotate around
the internuclear axis
and so $(J^1_0-L^1_0-S^1_0)=0$ (where the 0 component of the rotational angular momentum
operator is along the internuclear
axis due to the selection of the frame $F'$).

When the molecule is under the action of a magnetic field $\mathbf{B}$, the
magnetic sublevels pertaining to each energy level of
total angular momentum $J$ have slightly different energies, essentially due to
the
precession of the total angular
momentum around the quantization $z$-axis, which
we have chosen along the magnetic field vector. 
It is possible to write a Zeeman effective Hamiltonian for a diatomic molecule
that takes into account
the coupling between the magnetic field $\mathbf{B}$ and the total spin
$\mathbf{S}$ and orbital angular momentum 
$\mathbf{L}$ and between the magnetic field and the rotation of the molecule
$\mathbf{N}=\mathbf{J}-\mathbf{S}$:
\begin{eqnarray}\label{eq_5_26}
H_\mathrm{Z} &=& g_S \mu_0 \mathbf{B} \cdot \mathbf{S} + g_L \mu_0 \mathbf{B}
\cdot \mathbf{L} \nonumber \\
&-&g_r \mu_0 \mathbf{B} \cdot (\mathbf{J}-\mathbf{S}),
\end{eqnarray}
where $\mu_0$ is the Bohr magneton, $B$ is the magnetic field vector, $g_S$ is
the electron spin $g$-factor, $g_L$ is the
electron orbital $g$-factor and $g_r$ is the rotational $g$-factor (including
the nuclear and electronic contribution
$g_r=g_r^\mathrm{nuclear}-g_r^\mathrm{elec}$). If the spin of the nuclei is non-zero, a contribution
of the coupling between the nuclear spins and the magnetic field can be included with:
\begin{equation}\label{eq_5_26b}
H_\mathrm{Z}^{\mathrm{nuc}} = \sum_{\i} g_N^i \mu_N \mathbf{B} \cdot \mathbf{I}^i,
\end{equation}
where the nuclei are labeled by $i$. Because we neglect hyperfine structure, we will also neglect this term.

Although small differences exist, it is usual to put $g_S=2$ and $g_L=1$. The
exact value of $g_S$ is slightly 
larger than 2 (its value can be obtained from Quantum Electrodynamical
considerations) and in fact the values of the
$g$-factors can be slightly different due to the inclusion of additional
perturbations from other electronic states.
On the contrary,
due to the difference of mass between the electron and the proton, the
rotational $g$-factor (which is of the order of $\mu_N/\mu_0$, with $\mu_N$ the
nuclear
magneton) is $\sim$3 orders of magnitude smaller than the electronic
$g$-factors. Therefore, the contribution of the last term of the Zeeman
Hamiltonian
is only relevant when the molecule has no spin and no orbital angular momentum.
In this case, there is no
contribution to the Zeeman effect coming from the electronic motion because the
electronic cloud is 
essentially spherical and the molecule is in a $^1\Sigma$ state. This is the
case of the fundamental 
electronic states of CO and SiO, for which the sensitivity to the presence of a
magnetic field is very small.
Ideally, the value of the $g$-factors should be determined by
confronting the theoretical description of the molecule with spectroscopic
observations. The
rotational $g$-factor includes a pure nuclear contribution and another one
coming from the coupling between the rotation of the
electronic cloud and the magnetic field (\citarNP{judd75}). The nuclear
contribution to the rotational $g$-factor for a diatomic molecule 
formed by two nuclei A and B can be estimated by the following
formula (\citarNP{judd75}):
\begin{equation}\label{eq_5_28}
g_r^{\mathrm{nuclear}} \simeq \frac{\mu_N}{\mu_0} \frac{\left( Z_A M_B^2 + Z_B
M_A^2 \right)}{M_A M_B \left( M_A + M_B \right)},
\end{equation}
where $M_A$ and $M_B$ are the masses of each nuclei in atomic mass units and
$Z_A$ and $Z_B$ are the atomic number of each
nuclei. It is important to take into account that the importance of the rotation
of the electronic cloud can be decisive. For 
example, the calculated value for the nuclear contribution to the OH rotational
$g$-factor is 
$g_r^{\mathrm{nuclear}} \simeq 5.25 \times 10^{-4}$ while the measured value is
$g_r=-6.33 \times 10^{-4}$ (\citarNP{brown78}).

A simplification of the functional form of the Zeeman hamiltonian arises when the quantization axis 
$z$ lies along the magnetic field direction. In this case, the
scalar products between the magnetic field vector
and the angular momentum operators in Eq. (\ref{eq_5_26}) are transformed into
their corresponding
projections along this axis multiplied by the modulus of the magnetic field
vector.
These projections are the spherical components of the operators along the
$z$ axis of the laboratory frame:
\begin{equation}\label{eq_5_27}
H_\mathrm{Z} = g_S \mu_0 B \bar{S}^1_0 + g_L \mu_0 B \bar{L}^1_0 -
g_r \mu_0 B (\bar{J}^1_0-\bar{S}^1_0).
\end{equation}

\subsection{Matrix Elements}
The energy levels of a diatomic molecule in the presence of an arbitrary
magnetic field can be obtained via the numerical diagonalization
of the total effective Hamiltonian matrix. In contrast to previous approaches, ours
does not make use of any perturbational 
calculation since we diagonalize the full Hamiltonian including the Zeeman contribution. The first step is
to write
the effective Hamiltonian matrix in the chosen
Hund's case (a) basis set. The advantage of this basis set is that the majority
of the molecular
fixed frame components of the angular momentum operators are diagonal in this
basis, which largely simplifies 
the evaluation of the Hamiltonian matrix elements.
Of course, for problems where Hund's case (a) is a good representation of the
angular momentum
coupling of the particular rotational level of the electronic level under consideration, 
the Hamiltonian will be highly diagonal. In general, 
the Hamiltonian will have a non-diagonal
contribution that manifest deviations from the Hund's case (a) coupling.

One of the easiest ways to calculate the matrix elements of the effective
Hamiltonian 
is to substitute the rotational eigenfunctions by their
explicit expressions in terms of the rotation matrices given by Eq.
(\ref{eq_5_2}). The complex conjugate of the rotational
eigenfunctions $\langle J \Omega M|$ are obtained by taking the complex
conjugate of the rotation matrix and transforming them
with the help of the conjugation property given by Eq. (\ref{eq_5_3}). Since
many of the
matrix elements of the angular momentum
operators are known in the molecular fixed frame, it is
convenient to
evaluate them in this frame. In case the components of a tensor in the
laboratory
frame appear in the Hamiltonian (for example
in the Zeeman Hamiltonian), a transformation to the molecular fixed frame is
performed using the standard transformation 
rules for the tensorial components given by Eq. (\ref{eq_5_6b}). Finally, the
calculation of
the matrix elements of a spherical component of an arbitrary operator $A^k_q$ is
reduced to an integration over the Euler angles. The integrands of the $\langle J' \Omega'
M'|A^k_q| J \Omega M \rangle$ integrals consist
in the product of two or three rotation matrices (two from the rotational 
eigenfunction and an extra one from the transformation to the molecular fixed
frame). These integrals are easily calculated 
using the Weyl's theorem given by Eq. (\ref{eq_app_5_4}) and the property given by Eq.
(\ref{eq_5_5}) in Appendix
\ref{app_prop_rotation}. In other cases, it is advantageous
to directly use Wigner-Eckart's theorem given by Eq. (\ref{eq_5_7}). For facilitating the
calculation and
the reproducibility of our results, we give the
analytical expressions for the matrix elements of each 
term of the Hamiltonian in Appendix \ref{app_matrix_elements}.

\subsection{Diagonalization}
From an inspection of the explicit formulae for the matrix elements of the
effective Hamiltonian in Appendix 
\ref{app_matrix_elements},
we can see that many of the terms included for the description of the diatomic
molecule are diagonal in 
several quantum numbers.
For instance, the Hamiltonian associated to the spin-orbit coupling is
completely diagonal in Hund's
case (a) basis set. The molecular rotation contribution is diagonal in
$\Lambda$, $S$, $J$ and $M$, but not in\footnote{These are
good 
quantum numbers in the case of a non-rotating molecule, but they are not good
quantum numbers
when rotation is taken into account.} $\Sigma$ and $\Omega$, while the Zeeman interaction is diagonal
only in $\Lambda$, $S$
and $M$. Consequently, in the non-zero field case, the total angular momentum $J$ is
not a good quantum number. Therefore, 
only $M$ remains as a good quantum number and the effect of the magnetic field
has to be described in the
Paschen-Back regime. Strictly speaking, $J$ is not a good quantum number for any arbitrary value of the magnetic field strength.
Obviously, for sufficiently weak magnetic fields, $J$ is an approximately good quantum
number and we can safely treat the molecule in the Zeeman regime.

When selecting the basis set, we have to make sure that the chosen set of $|\Lambda S \Sigma; \Omega J M \rangle$ 
functions has to be complete enough to capture all the non-diagonal terms in the
Hamiltonian matrix. In our case, the Hund's
case (a) basis set was chosen using the following rules:
\begin{itemize}
\item We include eigenfunctions with the two possible values of the projection
of the orbital angular momentum on the internuclear
axis $\Lambda$, i.e., $\pm \Lambda$. This is fundamental to appropriately take
into account the
effect of $\Lambda$-doubling in which there is a break of the degeneracy between
energy levels with
the same value of $|\Lambda|$. Since in this paper we do not include any
$\Lambda$-doubling
interaction, the eigenvalues corresponding to both values of $\Lambda$ are
degenerate. However, if $\Lambda$-doubling needs to be explicitly accounted for, it is only a matter of
including the appropriate effective Hamiltonian and calculating its matrix
elements in Hund's case (a) basis set.

\item We include eigenfunctions with the two possible values of the projection
of the spin on the internuclear axis $\Sigma$, i.e., $\pm \Sigma$.

\item When calculating the energy of the magnetic sublevels of a given level
with total quantum number $J$, we also include the effect of the non-diagonal
terms between the $J$ level and the levels with $J-1$ and $J+1$. This is only
necessary when a magnetic field is
present because the contribution of the Zeeman splitting is
the only term included in the effective Hamiltoninan which is non-diagonal in
the quantum number $J$. In some cases (for instance, for heavy molecules), one may be interested
in including non-diagonal terms with $\Delta J=\pm 2$. With the present approach, this is only
a matter of augmenting the basis set by taking into account $J-2$, $J-1$, $J$, $J+1$ and $J+2$ for a
given level $J$.

\item For each value of $J$, we include all the possible values of $M$ from $-J$
to $J$. It is important to note that the Hamiltonian
is always diagonal in $M$ unless the hyperfine structure is taken into account.
\end{itemize}

The total number of eigenfunctions included in the basis set and, consequently,
the size of the 
effective Hamiltonian matrix, is obtained by summing all the possible
values of the quantum numbers: 2 possible values
of $\Lambda$ (1 in case $\Lambda=0$), $2S+1$ possible values of $\Sigma$ and 
$2J+1$ values of $M$ for each included value of $J$. Therefore, the size of the
basis set can be expressed in terms
of the spin of the electronic state and the $J$ value of the level of interest:

\begin{equation}
\label{eq_5_34a}
\mathcal{N} = \case{3(2S+1)(2J+1)}{\mathrm{if} \quad \Lambda=0, J \geq 1}{6(2S+1)(2J+1)}{\mathrm{if} \quad \Lambda \neq 0, J \geq 1.}
\end{equation}

For instance, let us assume we choose a $^2\Pi$ electronic state. The
eigenfunctions of the form
$|\Lambda S \Sigma; \Omega J M \rangle$
used for 
the calculation of the energies of a rotational level $J$ in this state 
are $|1,1/2,1/2;3/2,J',M \rangle$, $|1,1/2,-1/2;1/2,J',M \rangle$, 
$|-1,1/2,1/2;-1/2,J',M \rangle$ and 
$|-1,1/2,-1/2;-3/2,J',M \rangle$ (we have separated each number with commas for
clarity reasons). Since we include the 
coupling for the values
of $J$ between $J-1$ and $J+1$, we have to build such a set of eigenfunctions
for each value 
of $J'=J-1,J,J+1$. Additionally, we take into account the 
values of $M$ for each allowed value of $J'$ that are in the range
$-J'$ to $J'$. Summing up all the eigenfunctions obtained from the previous
combinations, we
end up with $24J+12$ eigenfunctions, a number which can be quite
high even for small
values of $J$.

\begin{figure}
\plottwo{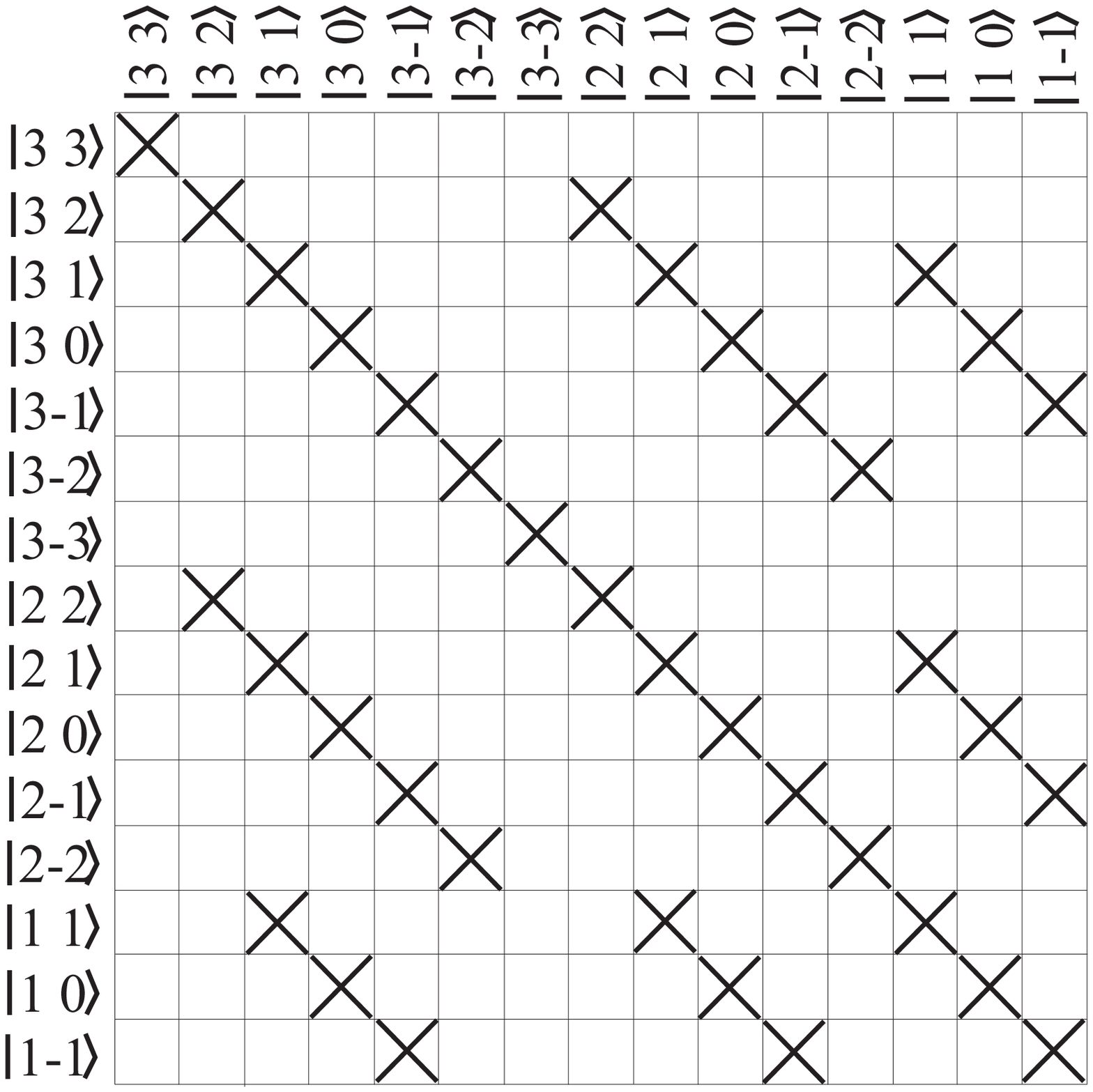}{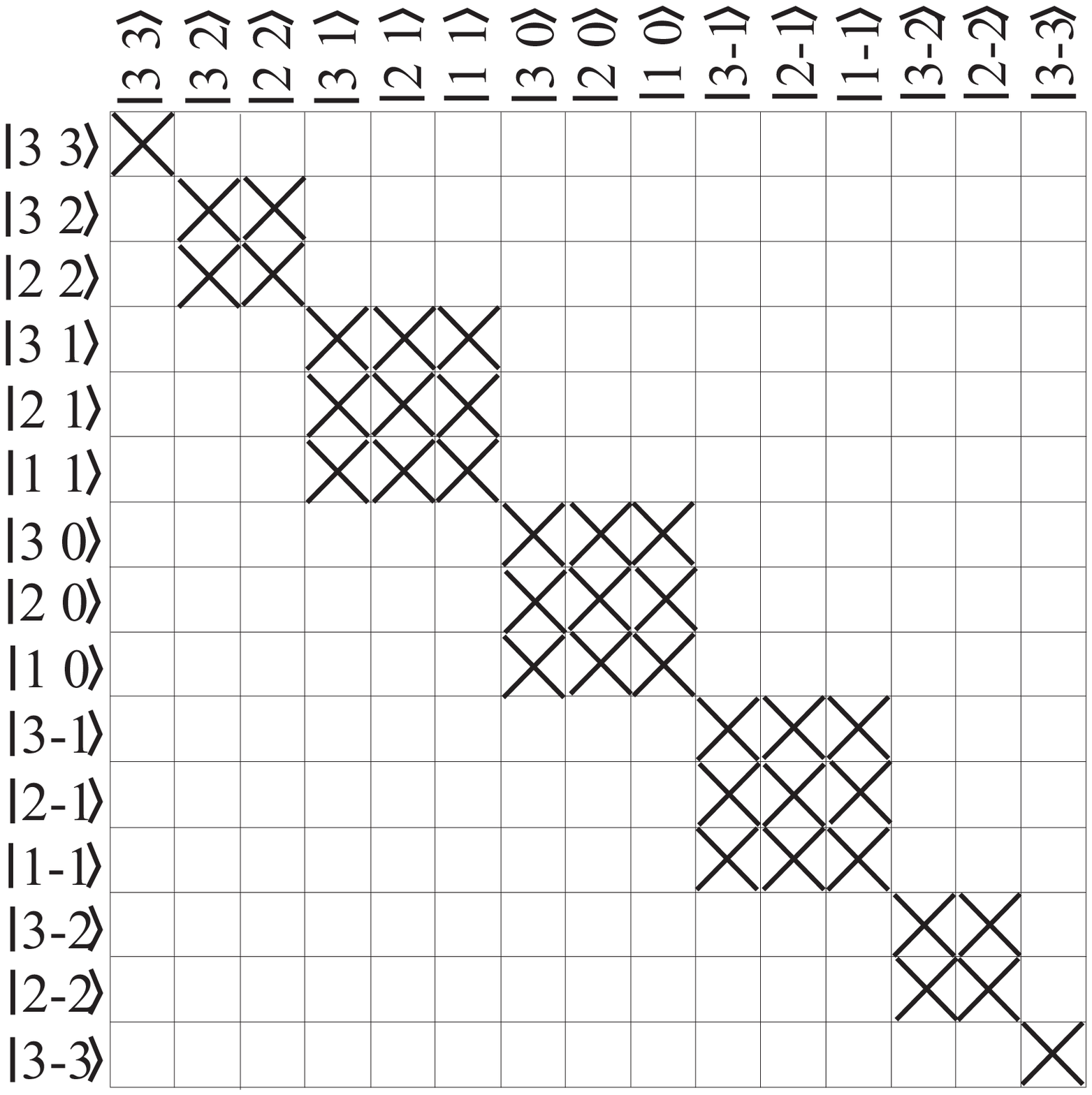}
\caption{Rotational part of the effective Hamiltonian matrix for a level with
$J=2$. A similar matrix has to be built for any combination of
$\Lambda$ and $\Sigma$. We indicate with a ``$\times$'' the elements which can,
in
principle, be different from zero. The rest of matrix elements are zero due to
the orthonormality of the eigenfunctions. Since we are indicating
only the rotational part of the Hamiltonian, the eigenfunctions are 
of the form $|J M \rangle$, with $\Omega$ prescribed by the values of $\Lambda$
and $\Sigma$. The left panel shows 
the Hamiltonian matrix when the basis set
is sorted by the value of $J$, 
that results in a highly non-diagonal matrix. The
right panel shows the same Hamiltonian matrix when the basis
set is ordered by the value of $M$. Since the effective
Hamiltonian is diagonal
in the $M$ quantum number, the resulting matrix is 
block-diagonal, thus allowing a block diagonalization in each of the
subspaces spanned by the eigenfunctions with the same value of $M$.}
\label{fig_hamiltonians}
\end{figure}

The Hamiltonian matrix built with these eigenfunctions is a $\mathcal{N} \times \mathcal{N}$
matrix. Due to the orthogonality properties of the eigenfunctions of Hund's case
(a), many of the 
matrix elements are zero. As a result, the
Hamiltonian matrix is very sparse (the number 
of non-zero elements is usually much less than the number of zero elements). In
order to numerically diagonalize this matrix, we can make use of any of the
available numerical procedures (see, e.g., \citarNP{numerical_recipes86}),
although algorithms specifically built for 
sparse matrices should be used in this case. As stated in Eq. (\ref{eq_5_34a}),
the size of the 
matrix increases linearly with the value of $J$ and the calculation of the
eigenvalues and
eigenvectors of a big matrix represents
a very hard numerical work even for intermediate values of $J$. However, we have
seen that
the effective Hamiltonian matrix is diagonal in the subspaces spanned by
eigenfunctions with different 
values of $M$ (or $M_F$ in case we include the hyperfine structure). In this
case, if we reorganize the matrix
by ordering the basis set by their $M$ quantum number, we end up with a
block-diagonal matrix. Each block
belongs to the space spanned by the eigenfunctions with a given value of $M$.
Therefore, we transform the problem of diagonalizing a $\mathcal{N}\times \mathcal{N}$ matrix to the
diagonalization of $2J+1$ submatrices of 
size $6(2S+1)$. This
is shown in Figure \ref{fig_hamiltonians}, which corresponds to an
example with $J=2$. The standard algorithms applied to calculate the eigenvalues
of a general matrix scale as $\mathcal{O}(\mathcal{N}^3)$, so that this reorganization
translates into a huge decrease in the computing
time (of the order of $(2J+1)^2$).

With the numerical diagonalization procedure, we calculate the eigenvalues and
eigenvectors of the 
effective Hamiltonian matrix. The eigenvalues are associated to the energies of
the magnetic
sublevels $M$ of a given rotational level of the electronic state of interest.
The energy shift produced
by the presence of a magnetic field can be obtained by the difference between
these energies and the
zero-field energies, obtained by diagonalization of the effective Hamiltonian
neglecting the Zeeman 
Hamiltonian $H_Z$. The eigenvectors can be used then for calculating the
expectation value of any operator.
In our case, we are interested in calculating the expectation value of the
dipolar moment operator, that
is related to the relative strength of the transitions. The eigenfunction
associated with any level can
be written as the following linear combination of the eigenfunctions of Hund's
case (a):
\begin{equation}
| \Psi \rangle = \sum_{\Lambda \Sigma \Omega J} c_{\Lambda \Sigma \Omega J}
|\Lambda S \Sigma \Omega J M \rangle,
\end{equation}
where the sum is extended over the values of $\Lambda$, $\Sigma$, $\Omega$ and
$J$ included in the basis set. Since the
matrix elements of the dipolar moment operator are known in Hund's case (a) (see, e.g., \citarNP{schadee78}), it
is possible to use the previous linear combination
to calculate it in the new basis.

\section{Illustrative Results}
This section is devoted to showing some comparisons between the results obtained
via our numerical diagonalization of the Hamiltonian and with the previous
approach to the molecular Zeeman effect
based on Shadee's (1978) theory. Although the formalism of \citar{schadee78}
allows one to calculate the Zeeman splittings and
strengths of the Zeeman components of transitions among doublet states
($S=1/2$), it has been recently extended
to transitions among states with arbitrary spins but for the 
linear Zeeman regime
(\citarNP{berdyugina02a}). Our approach is suitable also for
calculating the Zeeman splittings and strengths of the components for the case
in which a transition to the Paschen-Back regime occurs
for electronic states with arbitrary spins. The
Hamiltonian diagonalization approach is the most general
one, since it permits us to include easily any 
additional coupling among the angular momenta of the molecule
and is valid for arbitrary strengths of the external
field. Moreover, it can be applied to any electronic state with an arbitrary
value of the spin.

Since one of the various applications we are carrying out
is the synthesis of Stokes profiles induced by the molecular
Zeeman effect in strongly magnetized regions of the solar
atmosphere, we show first the Zeeman patterns for several
lines of different diatomic species which are observed in the solar atmosphere.
As is well known, a
Zeeman pattern diagram shows the position and relative strength
of each of the $\sigma$ ($\Delta M = \pm 1$) and $\pi$ ($\Delta M=0$)
components which arise from transitions among the magnetic sublevels of the
upper and lower rotational levels of a given spectral line. The components in
the upper part of the diagram are the $\sigma$ components, with the $\Delta
M=-1$ transitions indicated as vertical lines
going upwards and the $\Delta M=1$ transitions indicated as vertical lines going
downwards. The relative strength of each component is proportional to the length
of the vertical
line. Additionally, the $\pi$ components are in the
lower part of the diagram. We have selected several molecular lines from MgH,
OH, CN, C$_2$, which show interesting Zeeman patterns
(see, e.g., \citarNP{schadee78}; \citarNP{illing81}; \citarNP{berdyugina02a};
\citarNP{asensio_trujillo_spw3_03}; \citarNP{asensio_trujillo05a}).
In addition, we show also the Zeeman patterns for the CCS radical, which is a
molecule of interest in the research field of star formation regions. Since this is a linear molecule,
it can be described with the same formalism as if it were a diatomic molecule.

\subsection{Doublet states}

\subsubsection{MgH}

Our first example is the P$_2$(5.5) line of the $A^2\Pi-X^2\Sigma^+$ (0,0) electronic
band of MgH. The ensuing MgH lines, which are located in the
visible range around  5150 \AA,
have recently become of interest because, apart from
presenting an interesting magnetic sensitivity, they show conspicuous scattering
polarization signals
when observed close to the solar limb
(\citartwoNP{stenflo_keller96}{stenflo_keller97}).

\begin{figure}
\plottwo{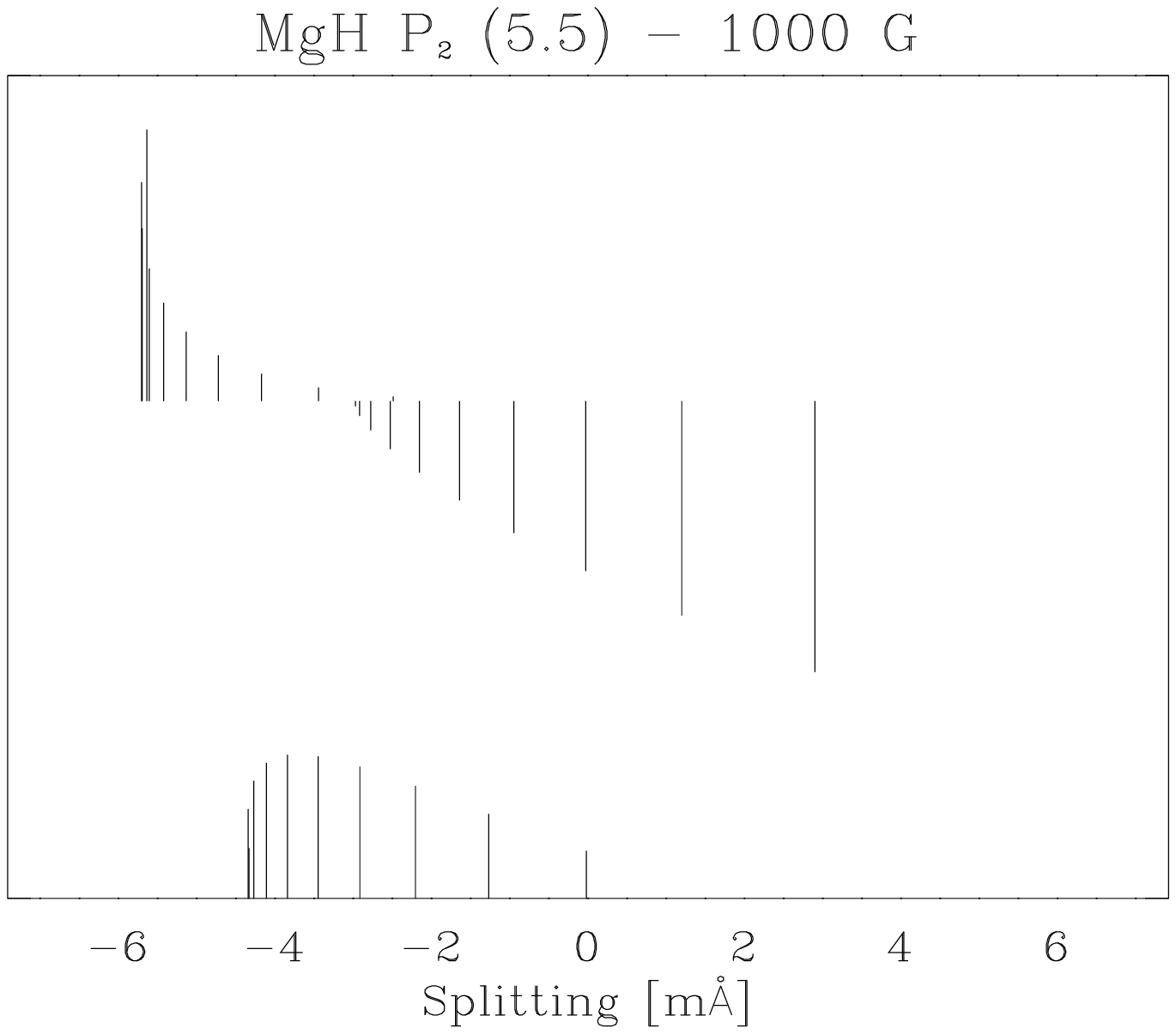}{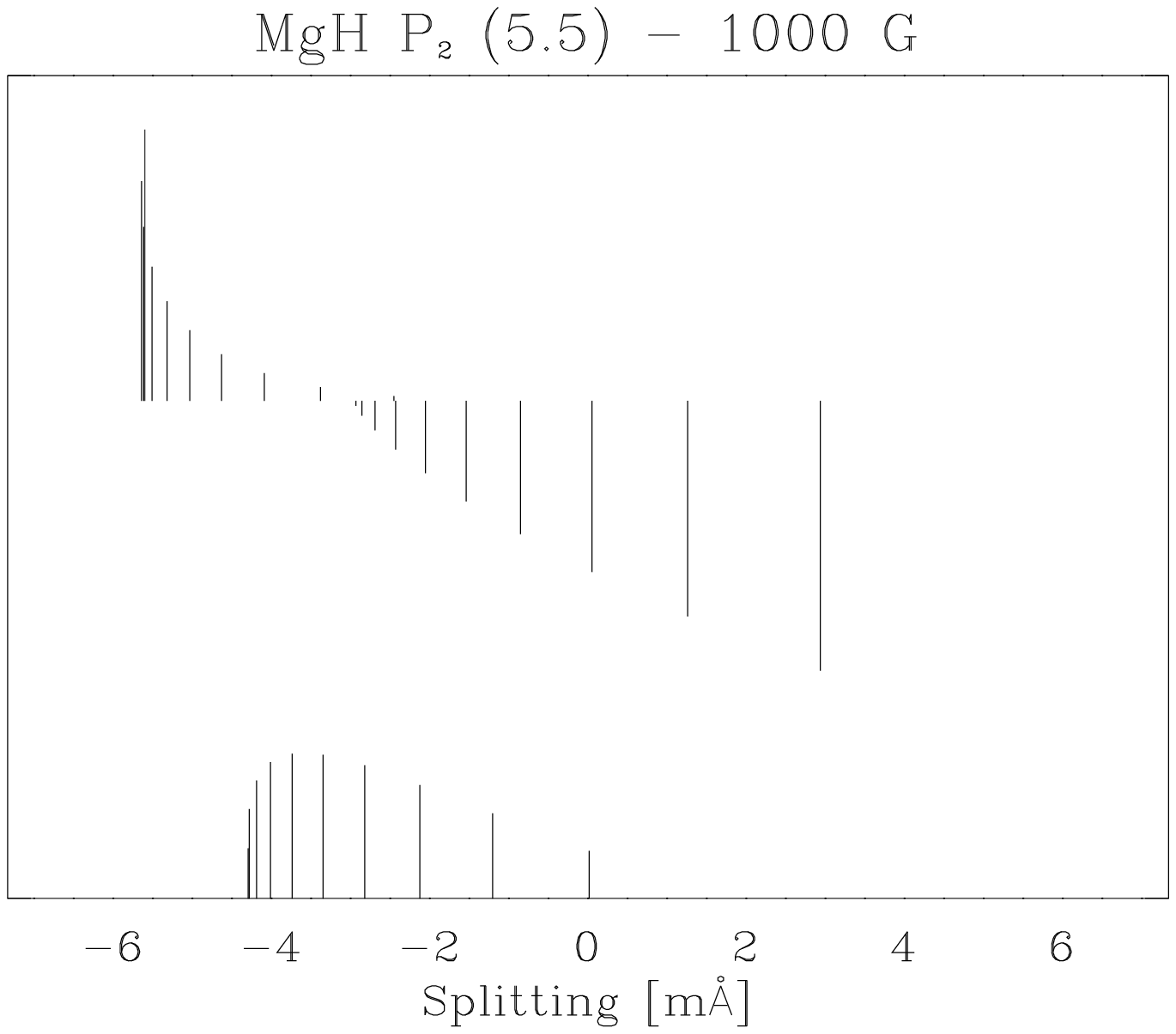}
\caption{Zeeman patterns for a magnetic field strength of 1000 G for the
P$_2$(5.5) line of the MgH electronic transition
$A^2\Pi-X^2\Sigma^+$.
The left panel shows the results applying the theory developed by
\citar{schadee78} while the right panel shows those obtained from the
numerical diagonalization of the Hamiltonian. This low-$J$ transition of MgH is
in the transition to the Paschen-Back regime for a field of
1000 G, thus the Zeeman patterns are highly perturbed due to the presence of the
non-diagonal $\Delta J=\pm 1$ terms in the total effective Hamiltonian.}
\label{fig_mgh_patterns}
\end{figure}

Figure \ref{fig_mgh_patterns} shows the Zeeman patterns obtained for a magnetic
field
strength of 1000 G. The left panel concerns the results obtained
using the theory developed by \citar{schadee78} while the right panel shows the
results obtained using our numerical diagonalization of the
Hamiltonian. The molecular constants have been obtained from
\citar{huber_herzberg03}. For consistency with the assumptions of
\citar{schadee78} and in order to properly
compare the results, we have included the very same terms in the effective
Hamiltonian.
In this case, the total Hamiltonian of the $A^2\Pi$ electronic state has only
contributions from the
spin-orbit coupling, while that of the lower $X^2\Sigma^+$ level has only
contributions from the spin-rotation coupling.
Figure \ref{fig_mgh_patterns} shows that both 
Zeeman patterns are indistinguishable, a proof
that our numerical
diagonalization technique is working properly. As previously indicated by
\citar{schadee78} (see also \citarNP{berdyugina02a}), such MgH lines present
strong Paschen-Back effects. The reason
is that the lower electronic state has a very small spin-rotation constant and
the energy gap between two
consecutive energy levels is very small. Therefore, the presence of a weak
magnetic field leads to a Zeeman splitting
comparable to this energy gap, which produces interferences 
between the magnetic substates. The presence
of Paschen-Back effects produce a deformation of the
Zeeman pattern with respect to the symmetric situation 
of the linear Zeeman regime. The ultimate cause of such interferences is
that the Zeeman Hamiltonian $H_Z$ is not diagonal
in $J$, so that the energy levels do not have a $J$ value associated with them.
When a magnetic 
field is present each energy level has contributions from levels having
different values of $J$ in the zero-field case.

\subsubsection{OH}

\begin{figure}
\plottwo{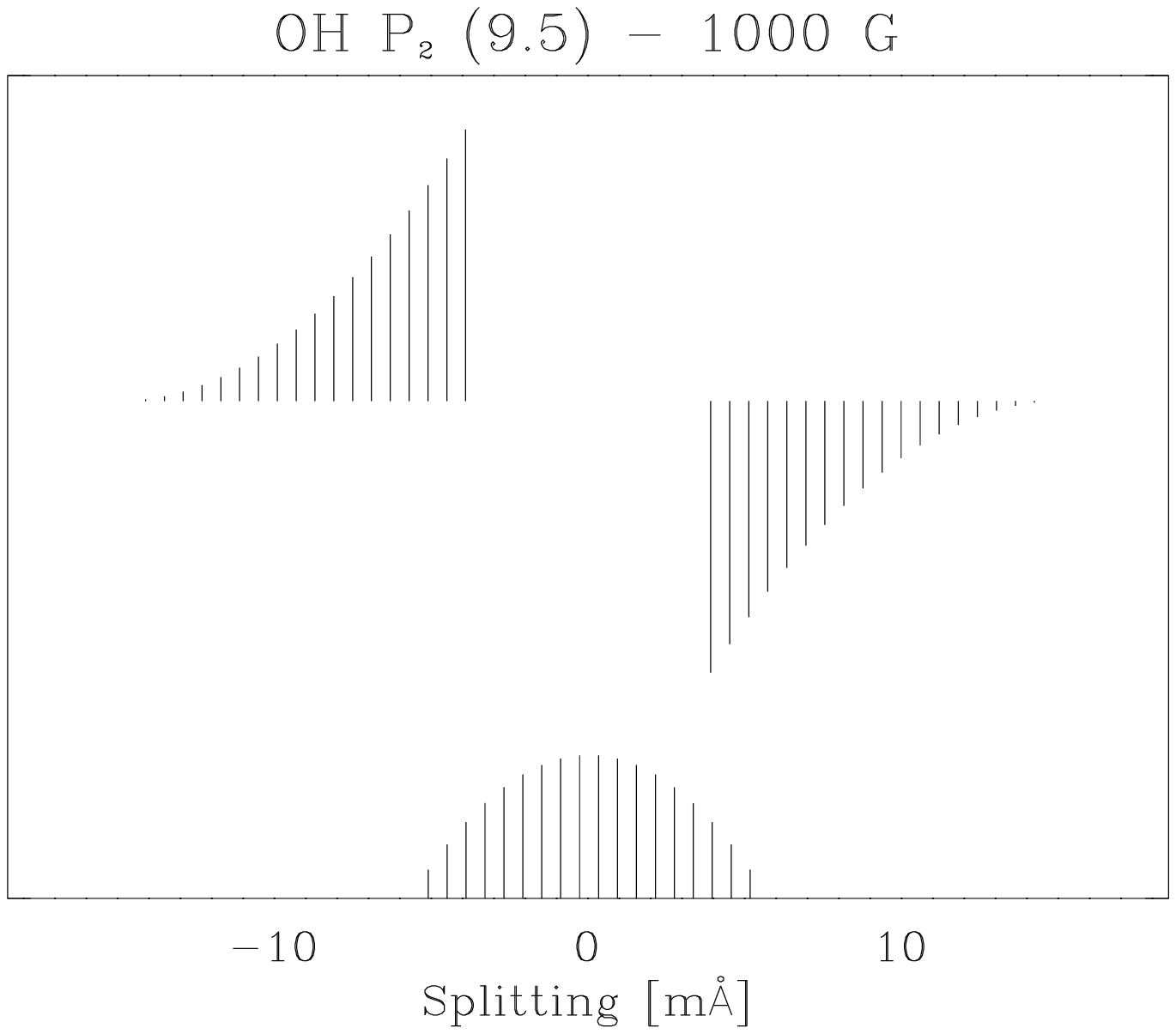}{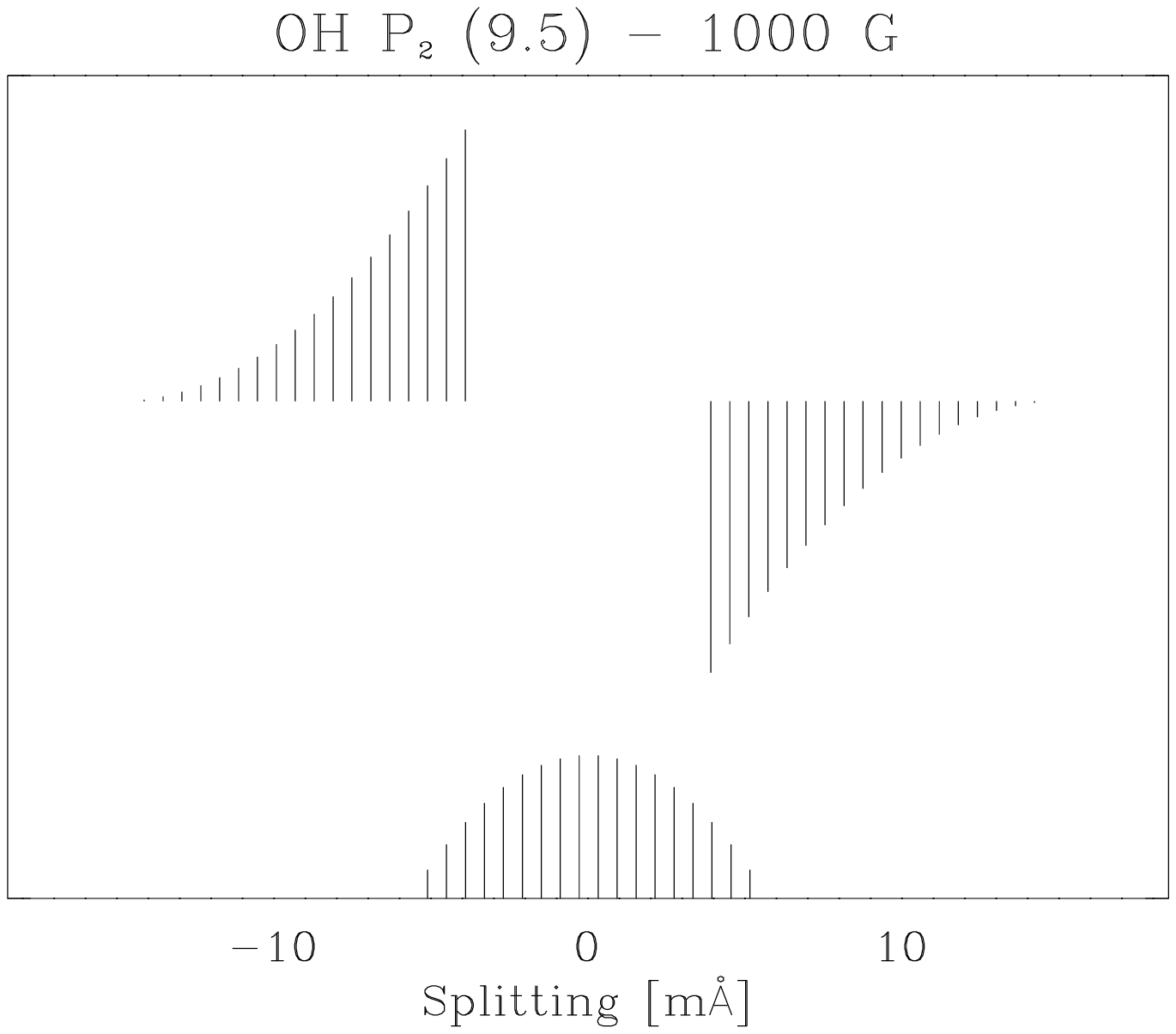}
\caption{Zeeman patterns for a magnetic field strength of 1000 G for the
P$_2$(9.5) line between vibrational levels of the fundamental
$X^2\Pi$ state of OH. The left panel shows the results applying the
theory developed by \citar{schadee78} while the right panel shows those obtained
from the numerical diagonalization of the
Hamiltonian. The magnetic field at which there are interactions between adjacent
rotational levels in this electronic state is much higher than
1000 G, so that this line is in the Zeeman regime. As a consequence, the Zeeman
patterns are almost unperturbed. However, the transition
has to be described using an intermediate coupling between Hund's cases (a) and
(b).}
\label{fig_oh1_patterns}
\end{figure}

\begin{figure}
\plottwo{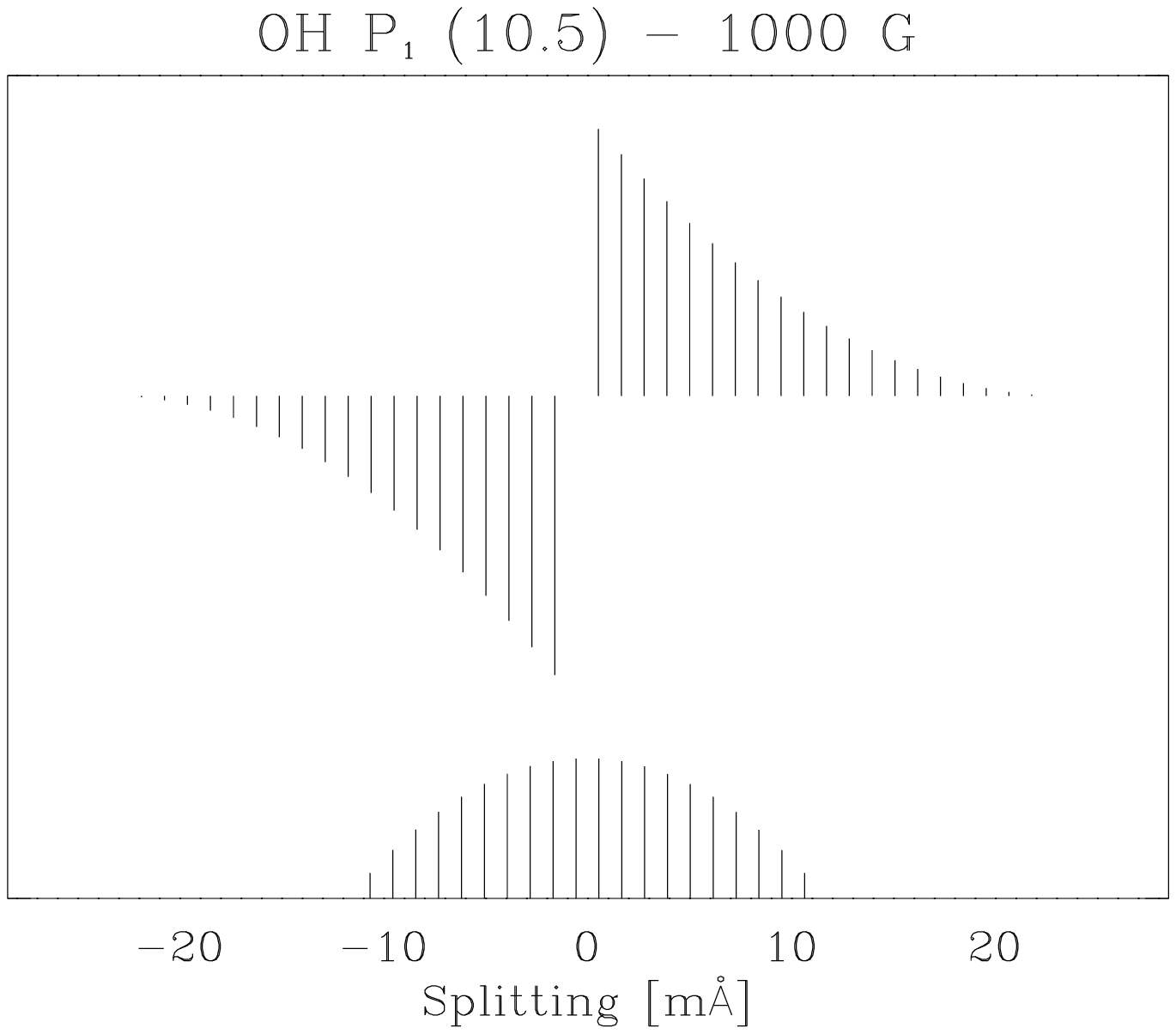}{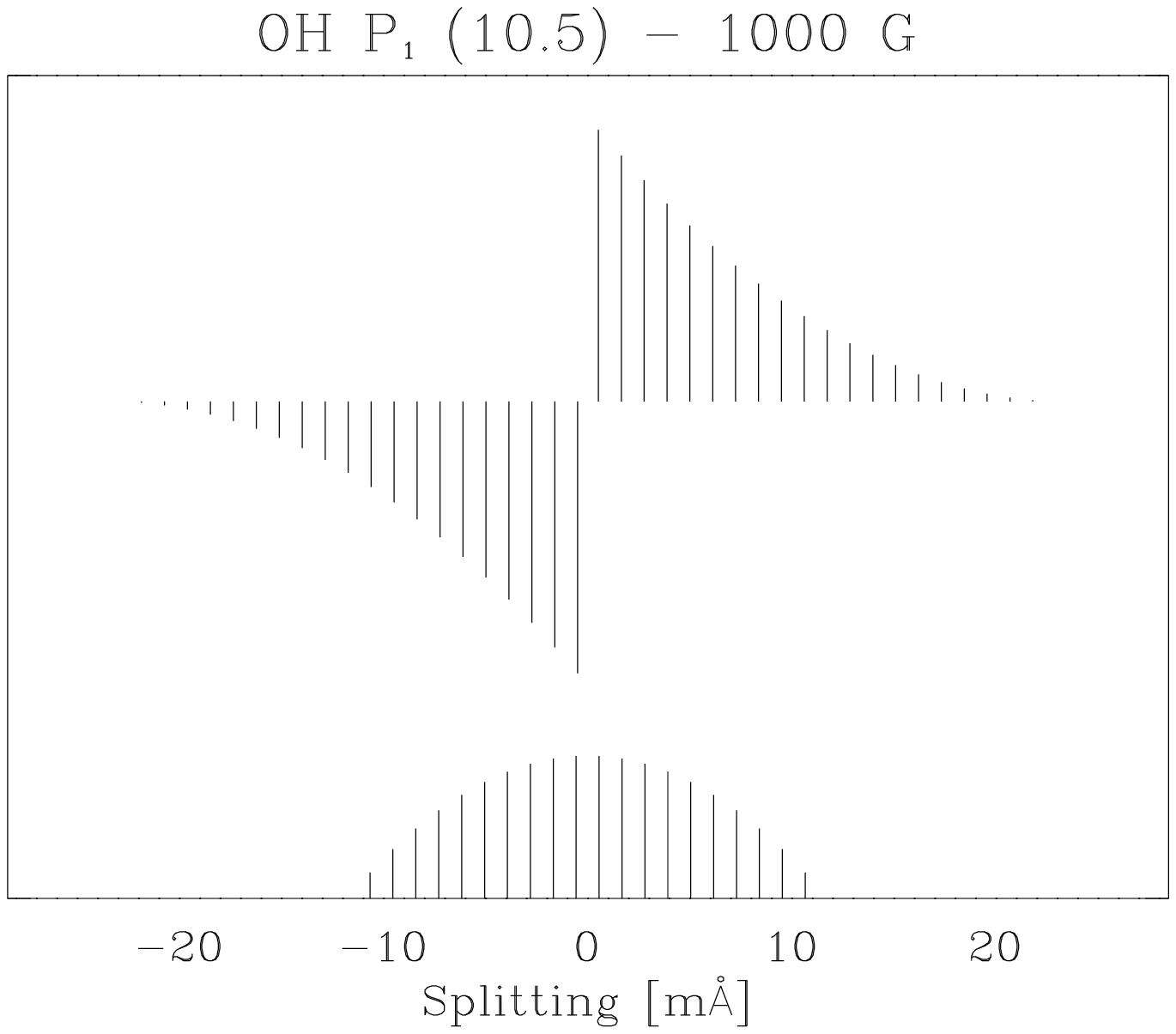}
\caption{Same as Figure \ref{fig_oh1_patterns} but for the P$_1$(10.5) line
between vibrational levels of the fundamental $X^2\Pi$
state of OH.}
\label{fig_oh2_patterns}
\end{figure}

The next two examples shown in Figures \ref{fig_oh1_patterns} and
\ref{fig_oh2_patterns} are for the P$_1$(10.5) and P$_2$(9.5) near-IR lines
of OH, which belong to the vibro-rotational (2,0) band of the fundamental electronic
state $X^2\Pi$. Here the transition to the
Paschen-Back regime occurs for magnetic strengths 
slightly below 10$^6$ G
(\citarNP{berdyugina02a}). Therefore, for magnetic fields of 
${\sim}1000$ G we are safely in the Zeeman regime. 
Similarly to the MgH case, the left panels show the results obtained with
the theory of \citar{schadee78}, while the right panels
show the results that we have obtained when using the numerical diagonalization
of the Hamiltonian
including only those terms which were considered by \citar{schadee78}. The
molecular constants have been also obtained from
\citar{huber_herzberg03}. Note the extremely good agreement between both
calculations. Since these OH lines are in an intermediate
coupling scheme between Hund's case (a) and (b)
(\citarNP{berdyugina02a}), this calculation
constitutes another proof that our numerical diagonalization
technique is working properly. 
Although we have written the Hamiltonian matrix
using the basis functions of Hund's case (a), we can correctly
recover the behavior for levels which are not correctly described under this
case. In this situation, the Hamiltonian is non-diagonal and
the final eigenfunctions result from appropriate combinations of the Hund's case
(a) eigenfunctions.

The circular polarization profiles produced by
these OH lines in sunspot umbrae were first observed by \citar{harvey85} and
modeled by \citar{berdyugina01}
taking into account that the effective Land\'e factor of the two lines have
opposite signs (\citarNP{ruedi95}).
The fact that both lines have effective Land\'e factors of opposite sign can be
noted in Figs. \ref{fig_oh1_patterns} and
\ref{fig_oh2_patterns}. To this end, we recall the definition of the effective
Land\'e factor as the wavelength shift from line center of the \emph{center of
gravity} of the $\Delta M=-1$ components in Lorentz
units (i.e., in units of $\lambda_B=\lambda_0^2 \nu_L /c$, where $\lambda_0$ is
the central wavelength, $c$ is the speed of light and
$\nu_L$ is the Larmor frequency of the magnetic field). The Zeeman
patterns show that the displacement from line center of the
$\Delta M=-1$ components have opposite signs for both lines.

With the polarization sensitivity we can obtain nowadays with the available solar polarimeters, the
effective Hamiltonian we have used appears to be sufficient for the description of the 
OH transitions. However, we must emphasize that additional
effects, like for instance the $\Lambda$-doubling, can be easily
included in case a more realistic description turns out to be
needed for explaining the emergent Stokes profiles.

\subsubsection{CN lines in the IR}

Another interesting example is the calculation of the Zeeman patterns for a
molecular transition with high $J$-values, since it permits to test the
numerical diagonalization approach for 
a case in which the effective Hamiltonian matrix is very large. One example is
the $Q_1(46.5)$ near-IR line of CN observed by
\citar{asensio_trujillo05a}. This line belongs to the $A^2\Pi-X^2\Sigma^+$
electronic transition of CN and it is located 
in the same spectral region as the
two OH lines discussed in the previous example.
The rotational constants were again taken from
\citar{huber_herzberg03}. The magnetic field at which the transition to the
Paschen-Back regime occurs for the levels of the upper $A^2\Pi$
electronic state is $\sim$560 kG (e.g., \citarNP{berdyugina02a}). Therefore, we
can safely treat them 
in the Zeeman regime for the typical stellar magnetic fields. However, the levels
of the $A^2\Pi$ electronic state have to be described in an intermediate
coupling scheme between Hund's cases (a) and (b).
On the other hand, the levels of the lower electronic 
state $X^2\Sigma^+$ can be
correctly described under the Hund's case (b) coupling
when the magnetic field is weak enough so that we are in the linear Zeeman
regime. However, the transition 
to the Paschen-Back regime in the lower electronic state occurs
for rather weak fields ($\sim$77 G for the lowest $J$
levels). For this reason, these lines are always in the Paschen-Back
regime under the typical magnetic fields of sunspots. 
This example represents then a complicated problem in which neither
the Zeeman regime nor any of the limiting Hund coupling cases
can be applied.

\begin{figure}
\plottwo{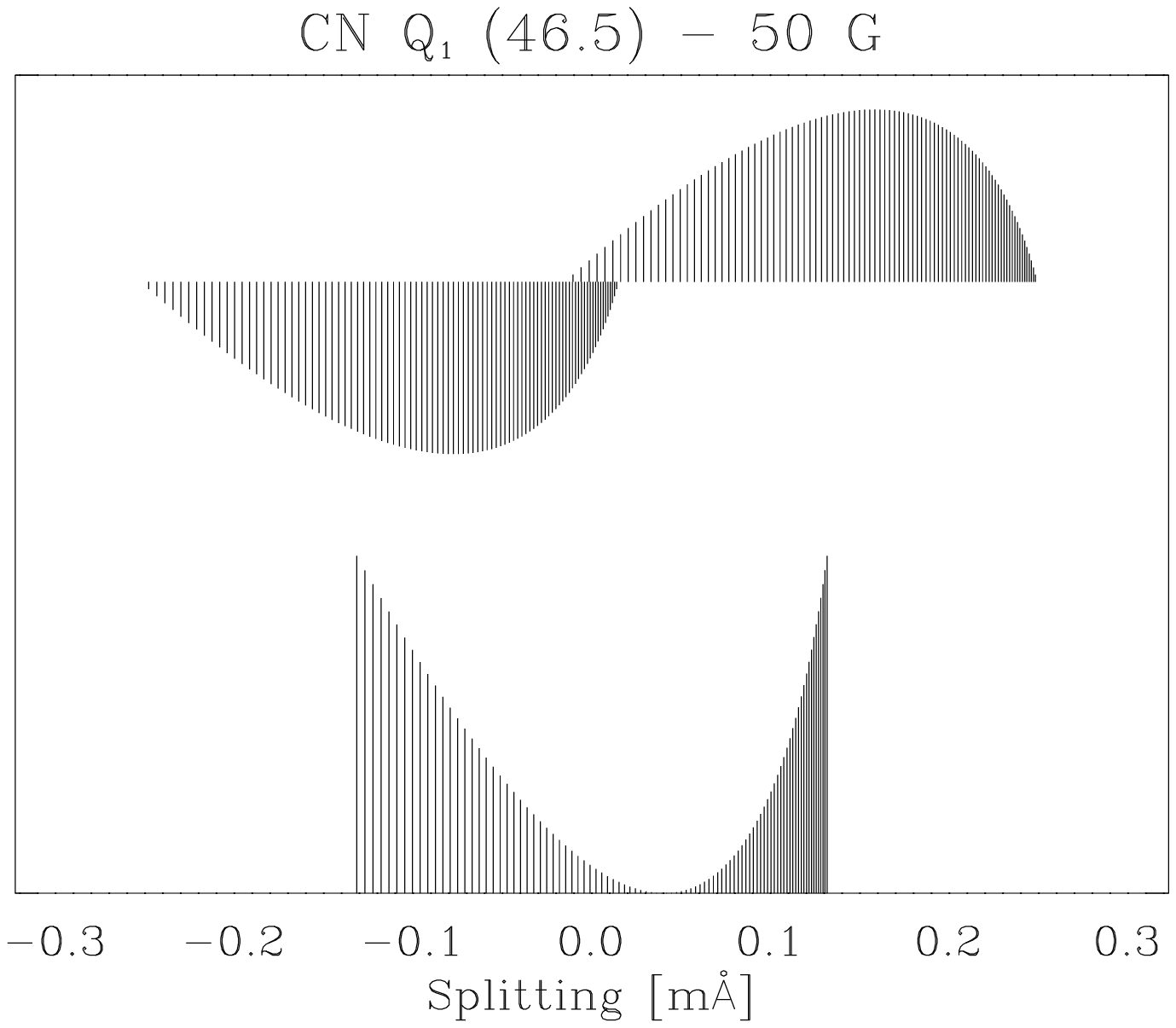}{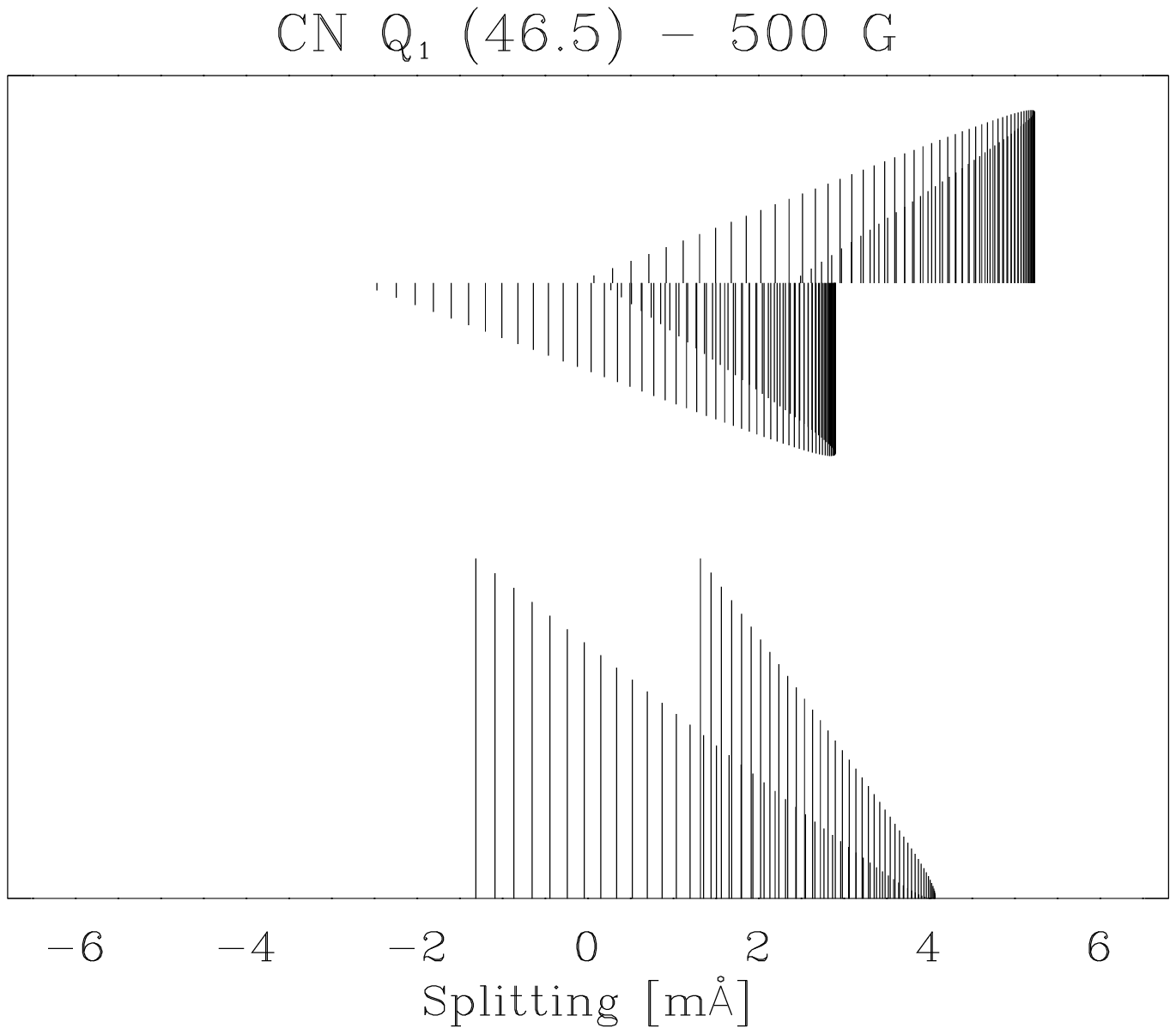}
\plottwo{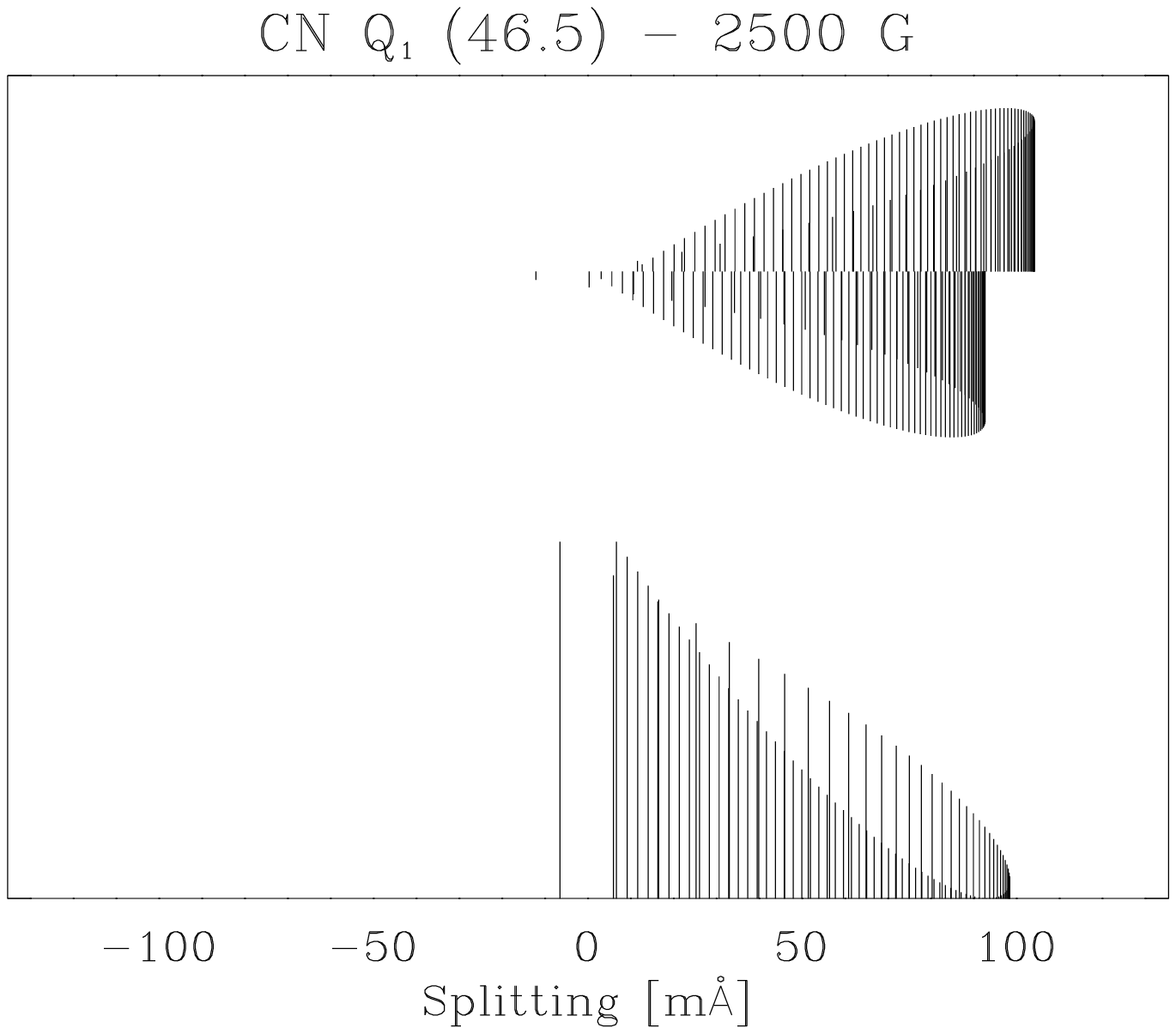}{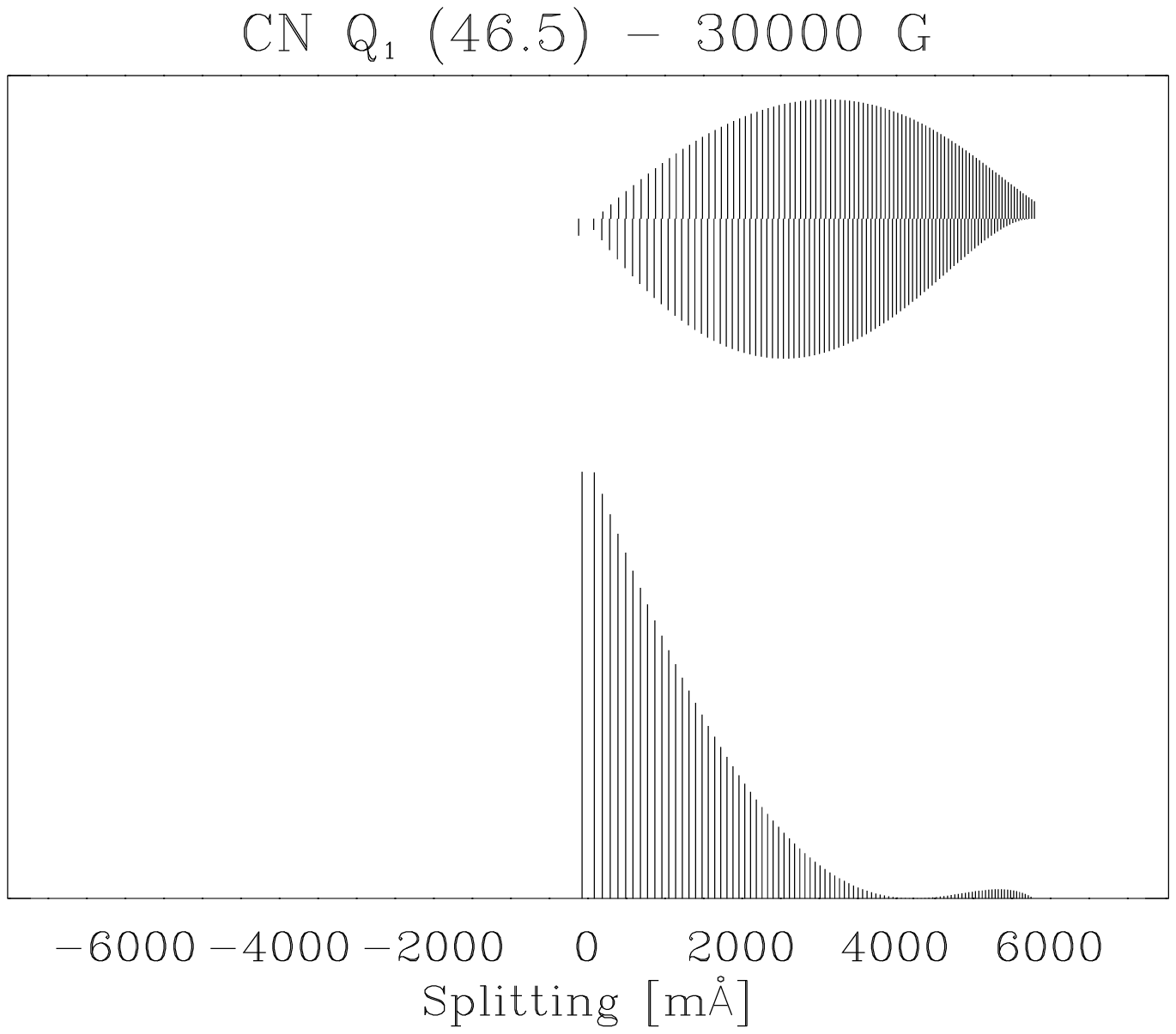}
\caption{Zeeman patterns for 50, 500, 2500 and 30000 G of the Q$_1$(46.5) CN
line. Note that the $\sigma$ and $\pi$ components
tend to be symmetric for low fields and get deformed due to the transition to the
Paschen-Back regime. At very high fields, the symmetry
is again recovered. This is an example of a transition in which the $\Delta J =
\pm 1$ matrix elements in the effective Hamiltonian are non-zero even for fields
as low
as 50 G.}
\label{fig_cn_patterns}
\end{figure}

\citar{asensio_trujillo05a} have pointed out that the 
observed polarization
signal from this and other
similar lines present an anomalous behavior, which is due to the peculiar form
of the Zeeman
patterns. In Fig. \ref{fig_cn_patterns} we show the Zeeman patterns for this
line
calculated via the Hamiltonian diagonalization technique. We have
verified that these results
are similar to those obtained using the formulation of \citar{schadee78}. Figure
\ref{fig_cn_patterns} shows the patterns for weak
fields (50 G), intermediate
fields (500 and 2500 G) and very strong fields (30000 G). The first one shows
that, even for weak fields, the CN lines cannot
be correctly described under the Zeeman regime and that interferences between
close $J$ levels are of importance. At the
typical kG field strengths of sunspots,
the $\Delta M=\pm 1$ components have almost the same center of gravity while
that of the $\Delta M=0$ components has a
displacement with respect to the $\Delta M=\pm 1$ components. This peculiarity,
produced by the Paschen-Back regime, leads
to the explanation of the anomalous emergent Stokes profiles in the umbral
spectrum.

\subsubsection{CN lines in the UV}

Another case of interest concerning transitions 
between doublet states is the
$B^2 \Sigma^+-X^2 \Sigma^+$ ultraviolet transition of CN.
This band has recently gained some attention due to the strong linear
polarization signals observed close to the solar limb
(\citarNP{stenflo_hawaii03}, \citarNP{gandorfer03}). Theoretical investigations
of the scattering polarization and the Hanle
effect in these molecular lines have been 
carried out by \citartwo{asensio_trujillo_spw3_03}{asensio_trujillo05c}
in order to explain the \emph{ladder}
structure of the observed $Q/I$ signals. Since the magnetic
field in the observed \emph{quiet} regions is assumed to
be relatively weak and the scattering polarization calculations are very
demanding, these
theoretical investigations were performed by using the Land\'e factors obtained
under Hund's case (b) coupling.

\begin{figure}
\plotone{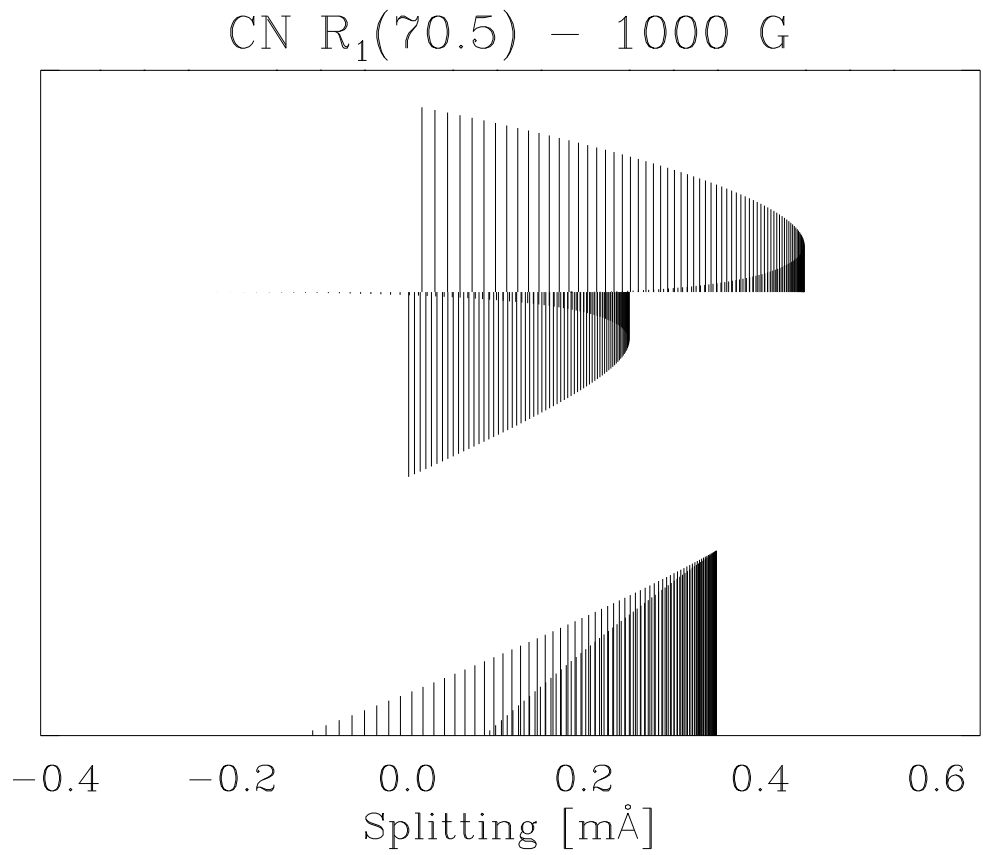}
\caption{Zeeman patterns for a magnetic field strength of 1000 G for the
R$_1$(70.5) line of the ultraviolet electronic transition
$B^2 \Sigma^+-X^2 \Sigma^+$ of CN. Note that the Paschen-Back effect is
present even for high $J$ values.}
\label{fig_cn_uv_patterns}
\end{figure}

Here we show that the Paschen-Back effect in this electronic system is of
importance for fields of the order of 1000 G (and even for fields below 100 G). The rotational constants were
taken from \citar{huber_herzberg03}. We have plotted only the results obtained
via the numerical diagonalization of the total Hamiltonian, but they
are similar to those obtained using the formulae of \citar{schadee78}.
Figure \ref{fig_cn_uv_patterns} shows the
Zeeman patterns for the 
$R_1$(70.5) line for a field strength of
1000 G. This result is representative of the
Paschen-Back effect for lines between high $J$ levels. The $R_1$(70.5) line is
one of the spectral
lines presenting strong scattering polarization signals in the spectral region
between 3771 \AA\ and 3775 \AA. It is interesting to note that Paschen-Back
effects 
are important for fields as low as 1000 G for transitions
between high- and low-$J$ levels. Therefore, for the 
typical kG fields of sunspot umbrae the lines of this system should present
anomalous Stokes profiles. The observable effects of these perturbed Zeeman
patterns on the 
emergent Stokes profiles should be similar 
to those observed in the lines of the 
$A^2\Pi-X^2\Sigma^+$ system, and whose Zeeman patterns
are shown in Figure \ref{fig_cn_patterns}.

\subsection{Non-doublet states}

\subsubsection{C$_2$}

\begin{figure}
\plottwo{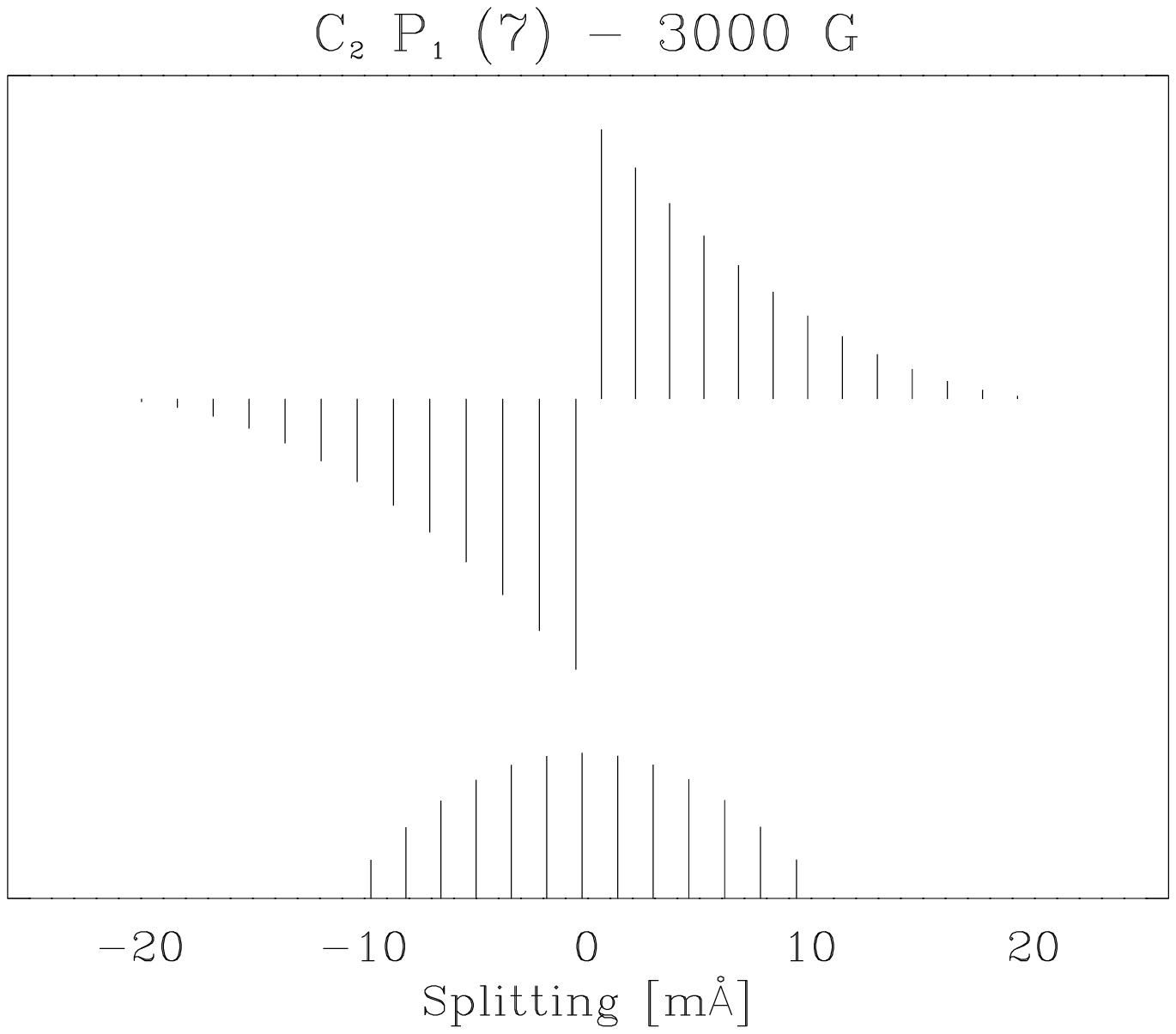}{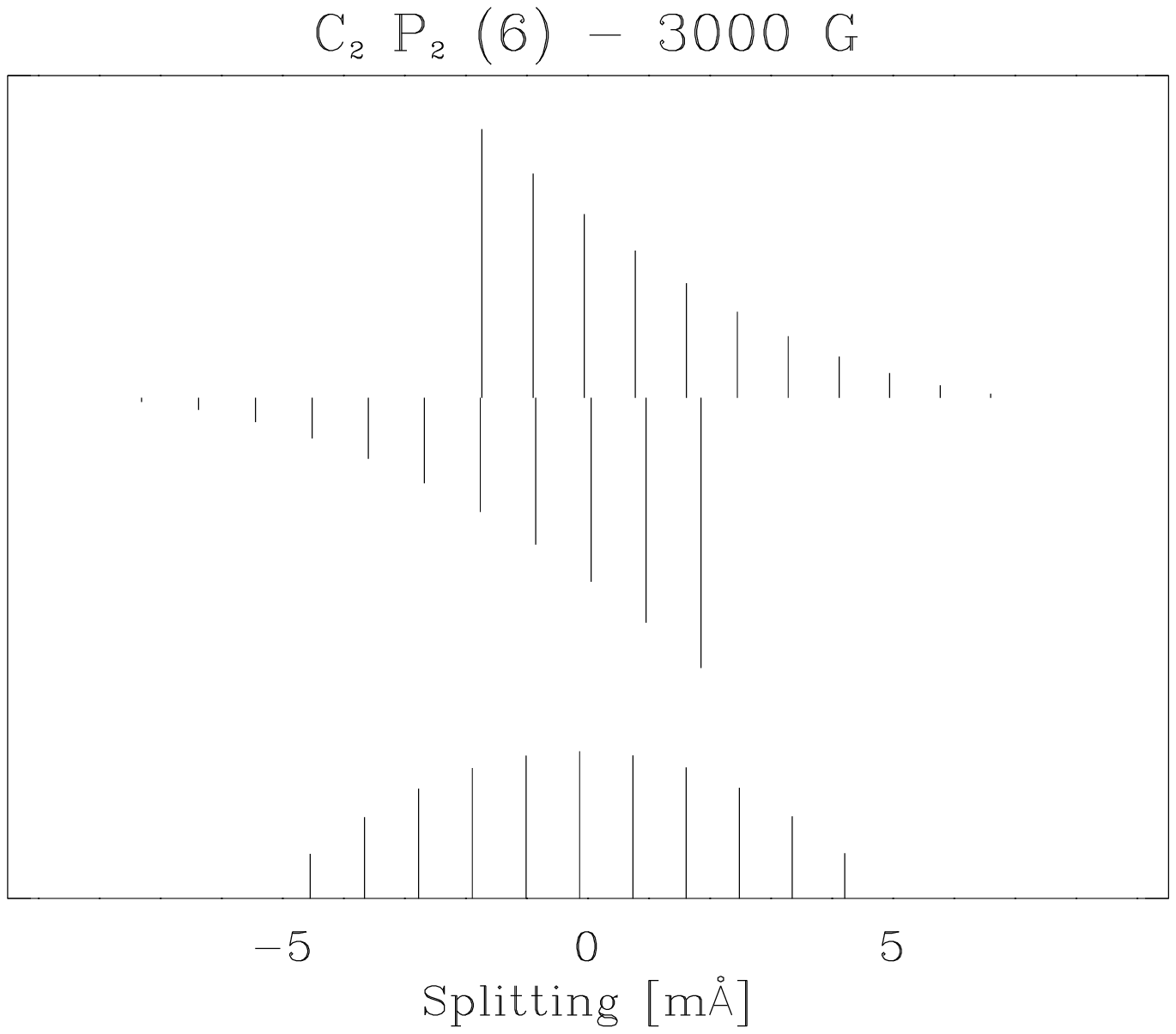}
\caption{Zeeman patterns for a magnetic field strength of 3000 G for the
P$_1$(7) and P$_2$(6) lines of the electronic transition
$d^3\Pi-a^3\Pi$ of C$_2$. In this case, since $S=1$, the theory of
\citar{schadee78} cannot be applied and these results have been obtained
with the numerical diagonalization of the Hamiltonian.}
\label{fig_c2_patterns}
\end{figure}

The first case we have considered is the (0,0) transition of the electronic
band $d^3\Pi-a^3\Pi$ of C$_2$; i.e., one of the so-called
Swan band. The investigation of this band in the solar atmosphere 
has gained some interest recently due to the linear
polarization signals produced by scattering processes
(see, e.g., \citarNPwiths{gandorfer_atlas1_00} atlas of the linearly-polarized solar limb
spectrum) and  their sensitivity to the presence of a weak magnetic field
via the Hanle effect (e.g., \citarNP{trujillo_nature04}). 
In this paper we are however
interested in calculating the Zeeman patterns of some lines of this 
band for a magnetic field of 3000 G. We have
chosen the P$_1$(7) and P$_2$(6) lines as representative of
the results which can be obtained also for other lines of the Swan system.
The results we have obtained via our numerical diagonalization of the total
Hamiltonian are shown in Figure \ref{fig_c2_patterns}.

The critical magnetic field at which one starts to detect signatures of the
transition to the Paschen-Back
regime for this electronic system is larger than
70000 G (\citarNP{berdyugina02a}). 
Since we are interested in much weaker
fields, the magnetic sensitivity of the 
lines of this transition can be safely modeled
in the Zeeman regime. This can also be inferred from the very symmetric
appearance of the Zeeman patterns shown in Figure \ref{fig_c2_patterns}. Furthermore, the
deviations from Hund's case (b) are not very large and we have verified that the obtained Zeeman patterns
have some similarities with those obtained in Hund's case (b). An
important conclusion regarding the effective
Land\'e factor can be obtained by a quick look to the 
Zeeman patterns. Note that the
center of gravity of the $\Delta M=-1$ transition is very
close to zero in the P$_2$(6) line while it is much larger for the P$_1$(7)
line. This shows that the P$_1$ lines
are much more magnetically sensitive than the P$_2$ lines. The same is valid for the P$_3$ lines.

It is very difficult to find molecules having electronic states
with $S \neq 1/2$ that present strong Paschen-Back effects
and which can be observed in solar and/or stellar spectra in the visible and
infrared regions. The target molecular transitions are
those in which the energy difference between consecutive levels is small enough
so that the Zeeman splitting for typical stellar magnetic fields 
produce significant interferences between
levels with consecutive values of $J$. Since the spin-orbit coupling is 
usually much larger than the spin-rotation one (see \citarNP{huber_herzberg03}
for the 
molecular constants), a candidate molecule should ideally have
an electronic state with $\Lambda=0$ so that the small spin-rotation
coupling gives strong interferences between closely lying energy levels. A few
molecules observed in the solar and other stellar atmospheres present
such states. This is the case of CO with its $a'^3\Sigma^+$ electronic state,
which is radiatively linked with the fundamental
$X^1\Sigma^+$ level by a forbidden transition around 1800 \AA. 

\begin{figure}
\plottwo{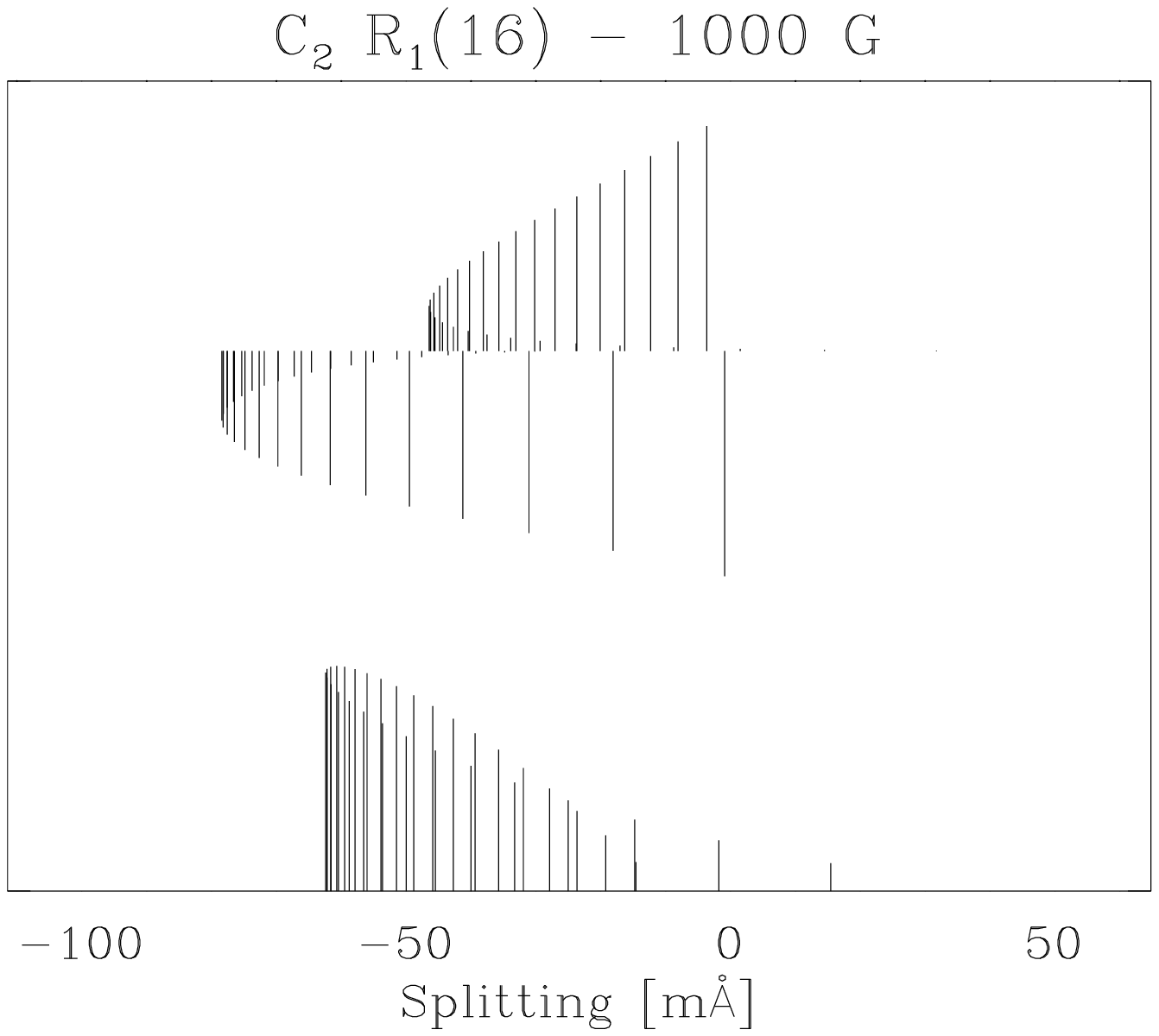}{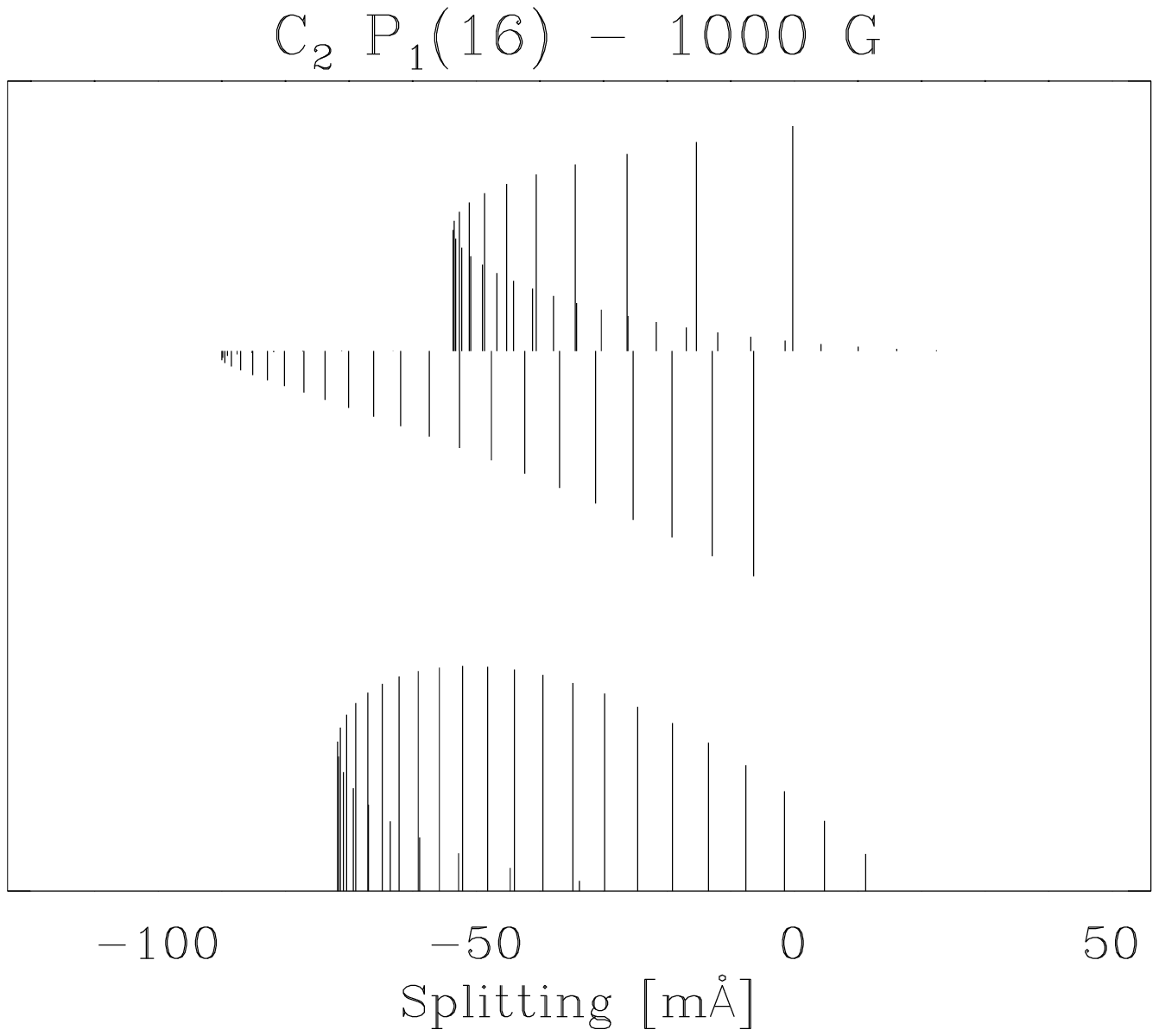}
\caption{Zeeman patterns for a magnetic field strength of 1000 G for the
R$_1$(16) and P$_1$(16) lines of the electronic transition
$b^3\Sigma^+_g-a^3\Pi_u$ of C$_2$ (the Ballik-Ramsay system). This is an example
of
a transition in which one of the levels has $S \neq 1/2$ while
having $\Lambda=0$. Since the energy separation of the multiplet levels is
small, the $\Delta J=\pm 1$ matrix elements of the effective
Hamiltonian are of importance because the splitting for moderate magnetic fields
are of the order of the energy separation. The Zeeman patterns
clearly show this perturbation in the transition to the Paschen-Back regime.}
\label{fig_c2_triplet_patterns}
\end{figure}

Another case of interest is
C$_2$, with the Ballik-Ramsay system. This system results from the electronic
transition $b^3\Sigma^-_g - a^3\Pi_u$ 
situated in the infrared (\citarNP{ballik_ramsay63}), 
with the center of the band around 1.8 $\mu$m. As an example, we
show in Figure \ref{fig_c2_triplet_patterns} the Zeeman patterns for the
R$_1$(16) and P$_1$(16) lines. Both lines
present strong Paschen-Back effects. Curiously, the deformed Zeeman
patterns resemble those patterns we calculated for the CN lines of
the $A^2\Pi-X^2\Sigma^+$ system. In fact, we expect that 
such C$_2$ lines present Stokes profiles that resemble those observed
in sunspot umbrae for the previously considered CN lines. Thus, they would present very weak
circular polarization signals and stronger and antisymmetric linear polarization
signals. Such C$_2$ transitions can be easily observed in carbon stars
(\citarNP{goorvitch90}) and they have been also detected in comets
(\citarNP{johnson83}). In this respect, it is of interest to mention that
\citar{johnson83} give a list of possible observable bands in comets. The
(0,0) band is around 5632.7 cm$^{-1}$,
the (1-0) band is around 7080.8 cm$^{-1}$ and the (2-0) band is around
8506.5 cm$^{-1}$.

\subsubsection{SO and CCS}

\begin{figure}
\plotone{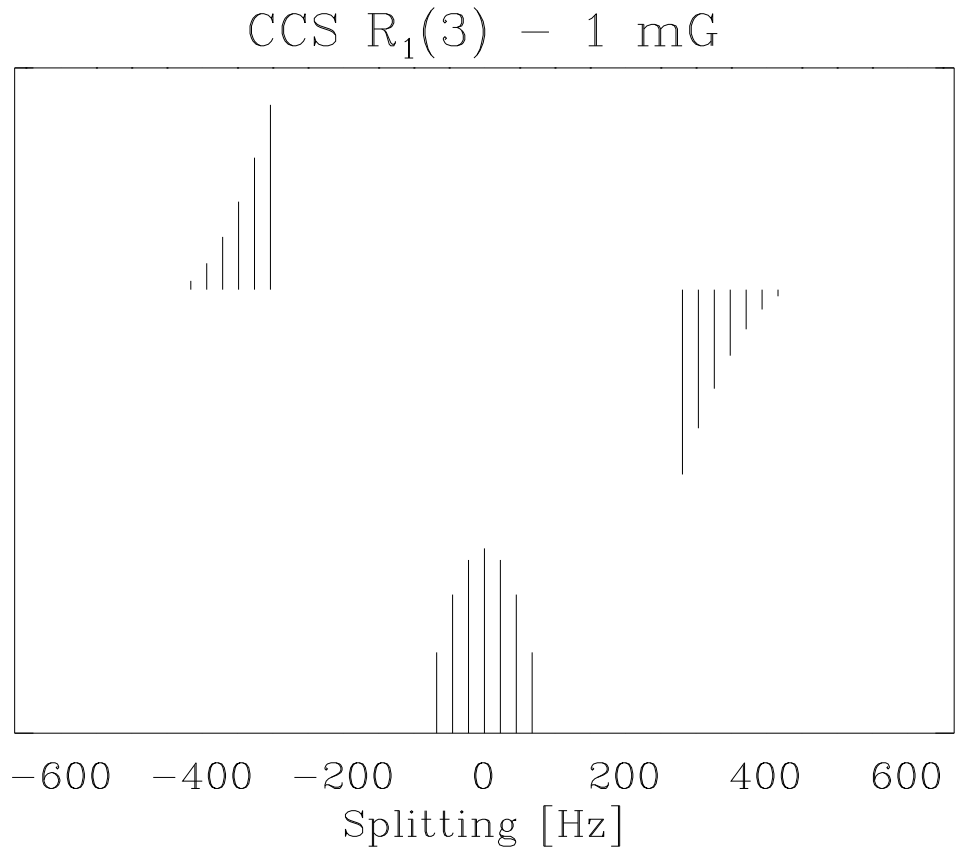}
\caption{Zeeman patterns for a magnetic field strength of 1 mG for the R$_1$(3)
pure rotational line of CCS in the fundamental 
$^3\Sigma^-$ electronic state. This line has to be described in an intermediate
coupling case between Hund's case (a) and (b). The comparison 
with the results of \citar{shinnaga_yamamoto00} indicate that our approach based
on a Hund's case (a) basis set works for transitions which
are almost perfectly described under a pure Hund's case (b) coupling.}
\label{fig_ccs_triplet_patterns}
\end{figure}

Turning now our attention to the millimeter-wave region, it is worthwhile to
mention that it is possible to observe pure rotational transitions
in the lower rotational levels of molecules
with ground electronic states having $\Lambda=0$ and $S \neq 1/2$. Some of
these
molecules are present in regions of star formation and the investigation of the
Zeeman effect
in these molecules is very important for gaining information about the magnetic
field. This is the case of SO and the CCS radical, which are used as tracers of
dense cores in star formation regions.
Since they present
no hyperfine structure, and due to the observed narrow and intense emission
peaks, they have been used to measure the velocity structure
in dense cores (\citarNP{langer95}). 
Due to the two unpaired electrons present in
both molecules, the Zeeman splittings are important, and this fact
has been used for searching for magnetic fields in the
densest parts of star formation regions.
\citar{shinnaga_yamamoto00} have calculated the Zeeman effect in the
$^3\Sigma^-$ fundamental electronic state of both SO and CCS in
order to give accurate Land\'e factors for several rotational lines.
Figure \ref{fig_ccs_triplet_patterns} demonstrates that we 
are able to reproduce
their Fig. 2 for the R$_1$(3) transition of CCS. Of course, we have used the
same rotational and coupling constants in order to be able 
to compare properly both results. 
The results of 
\citar{shinnaga_yamamoto00} were obtained by diagonalization of the effective
Hamiltonian
written using Hund's case (b) eigenfunctions. In this case, it is more
appropriate because, for weak rotation, the 
angular momenta coupling is closer to Hund's case (b). 
Remarkably, using our approach 
based on Hund's case (a) eigenfunctions, we obtain the same results, thus
demonstrating again that our approach is very robust. Interestingly, we find
that we need to use a magnetic field strength of 1 mG instead of 100 $\mu$G in
order to reproduce their results. Apparently, the magnetic field indicated in the labels of
Fig. 2 in \citar{shinnaga_yamamoto00} appears to be inconsistent with the results given in
the text, that are otherwise correct. Finally, it is important to note that, since the energy separation of rotational
levels decreases as the weight of the molecule increases, it might be possible that the $\Delta J=\pm 2$ matrix-elements are non-negligible
for a correct modeling of the Zeeman splitting in some molecular species.

\section{Conclusions}

The numerical approach presented in this paper allows us to calculate 
the effect of a magnetic field on the energy levels of
diatomic molecules for arbitrary values of the total electronic spin
and of the magnetic field strength (i.e., it is valid for states of any
multiplicity and for both the Zeeman and incomplete Paschen-Back regimes). 
The ensuing computer
program we have developed gives the splittings of the molecular energy
levels and the strengths of the individual Zeeman components (i.e., the Zeeman
patterns of molecular transitions), which is the basic information
we use in our radiative transfer code for modeling the emergent
Stokes parameters from magnetized stellar atmospheres.
It is based on an efficient numerical diagonalization
of the effective Hamiltonian, which gives the eigenvalues and eigenvectors
for each magnetic sublevel, from where we obtain the energies of the molecular levels and the expectation values
of the dipole moment operator between the pair of
levels producing the molecular transition under consideration.

Our numerical
diagonalization approach of the effective Hamiltonian can treat cases in which
quantum interferences
between very close in energy levels lead to the transition to the Paschen-Back
regime.
We take into account the effect of non-diagonal terms between levels with
$\Delta J= \pm 1$
in a self-consistent way, without making use of any perturbative calculation.
Our strategy 
generalizes previous ones which only allowed the investigation of 
doublet states for arbitrary values of the magnetic field (\citarNP{schadee78})
or states with arbitrary spin but in the linear Zeeman regime
(\citarNP{berdyugina02a}). Additionally, our computer program allows a
straightforward inclusion of any additional term into the effective Hamilonian,
so that the description of the molecular motion can be as refined as needed.

We have performed several comparisons 
between our results and those
obtained via previous formulations based on Schadee's (1978) theory,
demonstrating that both results are
indistinguishable when the very same 
terms of the effective Hamiltonian are included. We have
also performed calculations for several molecular transitions arising from
non-doublet states (e.g., states with $S=1$). Since the
theory developed by \citar{schadee78} is limited to doublet states, we have
performed comparisons in the Zeeman regime with the results
given by pure Hund's coupling cases. Furthermore, we have also 
calculated Zeeman patterns for diatomic lines in triplet states in which
clear Paschen-Back effects are at work.

It has been clearly shown in physics laboratory experiments that a very detailed description of the molecular motion is needed to
investigate correctly the magnetic properties of diatomic molecular lines (e.g.,
\citarNP{brown_carrington03}, and references
therein). 
However, in astrophysics the investigation and application of the Zeeman
and Hanle effects in molecular lines is still at an early stage of development. 
Fortunately, a new generation of polarimeters and telescopes
should allow us to obtain high-quality spectropolarimetric observations with
unprecedented spectral resolution and polarimetric sensitivity. We believe
that the numerical diagonalization approach described in this paper constitutes
the ideal framework to investigate in detail the magnetic properties of
molecular lines in different astrophysical environments, 
since it permits a rather straightforward inclusion
of all the desired terms in the effective Hamiltonian. Among the various
investigations we have in mind for the near future, we would like to mention
the influence of the hyperfine structure and/or the $\Lambda$-doubling on 
both the Zeeman and Hanle effects in the spectral lines of diatomic molecules.

\acknowledgments
We thank Egidio Landi Degl'Innocenti and the anonymous referee for their careful reading of our paper and for suggesting
some useful improvements.
This research has been funded by the
European Commission through the Solar
Magnetism Network (contract HPRN-CT-2002-00313) and by the Spanish Ministerio de Educaci\'on y Ciencia
through project AYA2004-05792.

\begin{table}[h]
\footnotesize
\centering
\begin{tabular}{|c|c|c|}
\hline
Hund's case & Case (a) & Case (b) \\ \hline \hline
Good quantum numbers & $\Lambda$, $S$, $J$, $\Sigma$ ($\Omega=\Lambda+\Sigma$) &
$\Lambda$, $S$, $J$, $N$ \\ \hline
Degeneracy (non-rotating) &
$ \casesmall{2}{\mathrm{if} \Lambda \neq 0}{1}{\mathrm{if} \Lambda = 0}$ &
$ \casesmall{2(2S+1)}{\mathrm{if} \Lambda \neq 0}{2S+1}{\mathrm{if} \Lambda = 0}$ \\ \hline
Degeneracy (rotating) &
$ \casesmall{2}{\mathrm{if} \Lambda \neq 0}{1}{\mathrm{if} \Lambda = 0}$ &
$ \casesmall{2}{\mathrm{if} \Lambda \neq 0}{1}{\mathrm{if} \Lambda = 0}$ \\ \hline
Values of $J$ & $|\Omega|, |\Omega|+1, |\Omega|+2, \ldots$ & $|N-S|,|N-S|+1,\ldots,N+S$
\\ \hline
Values of $N$ & Not defined & $\Lambda, \Lambda+1, \ldots$ \\ \hline
Conditions & $A \Lambda$ $\gg$ $BJ$ & $A \Lambda$ $\ll$ $BJ$ \\ \hline
\end{tabular}
\caption{Brief description of Hund's cases properties.}
\label{tab_good_numbers}
\end{table}

\appendix
\section{Hund's cases}
\label{app_hund_cases}
This Appendix describes briefly the most common angular momentum coupling
cases present in diatomic molecules.
All angular momentum vectors
of the molecule (electronic orbital $\mathbf{L}$, spin
angular momentum $\mathbf{S}$ and rotational
angular momentum $\mathbf{R}$) form a resultant total angular momentum which is
designated by $\mathbf{J}$. Hund's coupling cases are
distinguished by the strength of the coupling among all angular momenta present
in the molecule. 

\begin{figure}
\plottwo{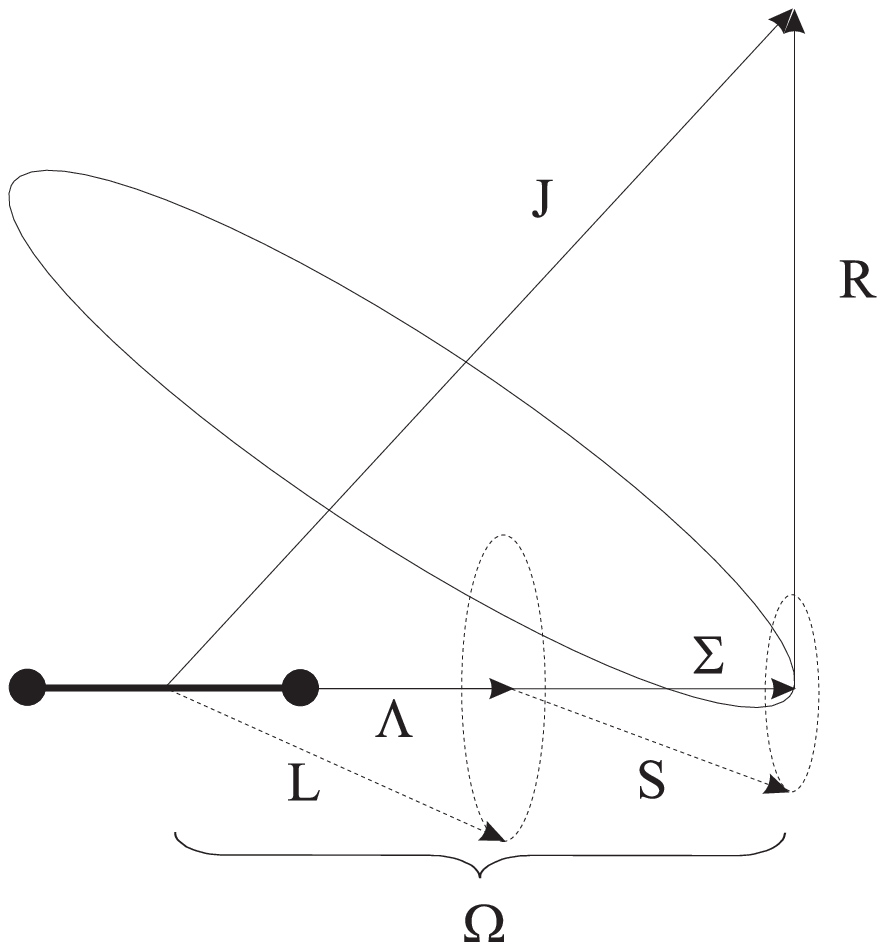}{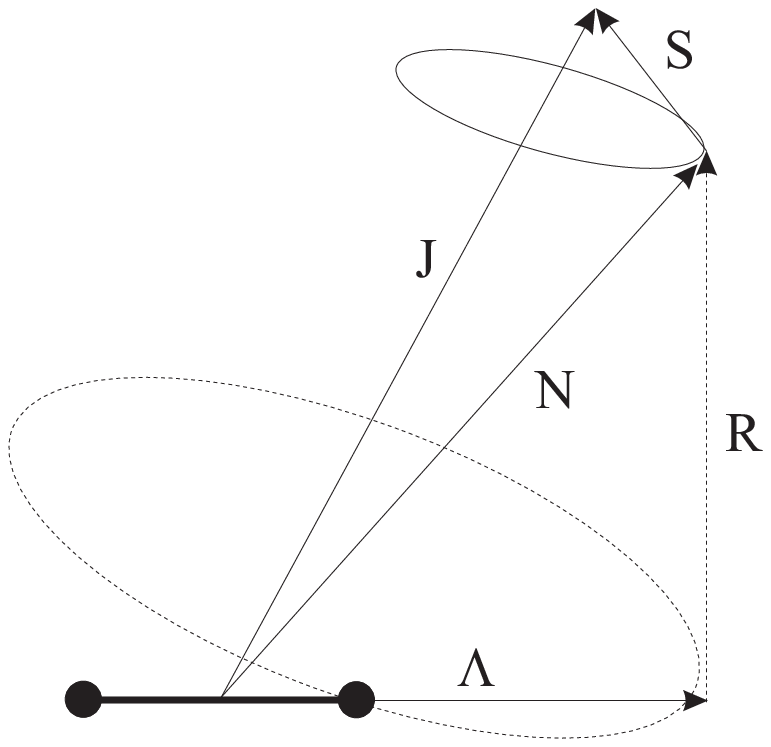}
\caption{Vector diagram for Hund's case (a) (left panel) and Hund's case (b)
(right panel). After \citar{herzberg50}.}
\label{fig_hunds_a}
\end{figure}

In Hund's case (a), illustrated in the left panel of Figure \ref{fig_hunds_a},
the orbital angular momentum is strongly
coupled to the internuclear axis by electrostatic forces. The spin angular
momentum is in turn strongly coupled to the
orbital angular momentum through spin-orbit coupling. Additionally, the
rotational motion is weakly interacting
with the spin and orbital motions. In this case, the projection of the total
electronic angular momentum
$\mathbf{\Omega}$ is well defined and composed of the projection of the orbital
angular momentum $\mathbf{L}$ on the internuclear
axis ($\Lambda$) plus the projection of the spin angular momentum
$\mathbf{S}$ on the internuclear axis ($\Sigma$).
The rotational angular momentum $\mathbf{R}$ thus couples with the angular
momentum along the internuclear axis $\mathbf{\Omega}$
to form the total angular momentum $\mathbf{J}$. The condition for 
the suitability of Hund's case (a) is that the spin-orbit
coupling $A \Lambda \Sigma$ has to be much larger than $BJ(J+1)$, with $B$ the rotational
constant and $A$ the spin-orbit coupling constant.

In Hund's case (b), illustrated in the right panel of Figure \ref{fig_hunds_a},
the spin angular momentum $\mathbf{S}$ is
very weakly coupled to the internuclear axis. In this case, the orbital angular
momentum $\mathbf{L}$ is still strongly
coupled to the internuclear axis with projection $\Lambda$. Because of the weak
coupling between the spin and the internuclear
axis, it is not possible to define $\Omega$. The vector $\mathbf{\Lambda}$
couples then to the rotational angular 
momentum $\mathbf{R}$ to form the resultant $\mathbf{N}$ (the total angular
momentum apart
from spin). Finally, the angular momentum $\mathbf{N}$ couples with the spin
$\mathbf{S}$ to form the total angular momentum
$\mathbf{J}$. The condition under which Hund's case (b) applies is that the
spin-orbit coupling must be much smaller than $BJ(J+1)$.

Table \ref{tab_good_numbers} gives a summary of the good quantum numbers in both
Hund's cases (a) and (b),
together with information on the degeneracy of the levels in the rotating and
non-rotating molecule. We also
give the possible values of $J$ and $N$ (if defined) and the conditions under
which each Hund's case applies.

\section{Useful properties of the rotation matrices}
\label{app_prop_rotation}
This appendix gives some important and useful properties of the rotation
matrices which can be found also in angular momentum books
(see, e.g., \citarNP{edmonds60}, \citarNP{brink_satchler68}, \citarNP{judd75}).
Two properties are used for deriving the
matrix elements of the Hamiltonian. The first one is the Weyl's
theorem, which states that the integral of the product of three rotational matrices
over the Euler angles can
be easily calculated as follows:
\begin{eqnarray}\label{eq_app_5_4}
\int_0^{2 \pi} d\alpha \int_0^{\pi} d\gamma \int_0^{2 \pi} d\beta \sin{\beta} \,
\rotmat{J_1}{M_1}{\Omega_1} (\alpha,\beta,\gamma)
\rotmat{J_2}{M_2}{\Omega_2} (\alpha,\beta,\gamma)
\rotmat{J_3}{M_3}{\Omega_3} (\alpha,\beta,\gamma) = \nonumber \\
= 8 \pi^2 \threej{J_1}{J_2}{J_3}{M_1}{M_2}{M_3}
\threej{J_1}{J_2}{J_3}{\Omega_1}{\Omega_2}{\Omega_3},
\end{eqnarray}
where we have introduced the 3-j symbols according to their standard definition (see, e.g., \citarNP{edmonds60}). The other property of
interest is
\begin{eqnarray}\label{eq_5_5}
\int_0^{2 \pi} d\alpha \int_0^{\pi} d\gamma \int_0^{2 \pi} d\beta \sin{\beta} \,
\rotmat{J_1}{M_1}{\Omega_1} (\alpha,\beta,\gamma)^*
\rotmat{J_2}{M_2}{\Omega_2} (\alpha,\beta,\gamma) = \nonumber \\
= \frac{8 \pi^2}{2J+1} \delta_{J_1 J_2} \delta_{M_1 M_2} \delta_{\Omega_1
\Omega_2},
\end{eqnarray}
where $\delta_{a b}$ is the Kronecker's delta, which is 1 for $a=b$ and zero
otherwise.

\section{Matrix elements of the effective Hamiltonian}
\label{app_matrix_elements}
This appendix gives the explicit form of the matrix elements of 
the effective Hamiltonian in Hund's case (a) basis set. One of 
the purposes is to show the 
explicit form of the matrix elements in order to recognize whether 
or not a contribution to the total
Hamiltonian is diagonal in Hund's case (a) basis set.
As it is mentioned in this paper, these expressions have been obtained by
applying the Wigner-Eckart
theorem to evaluate the tensorial form of the effective
Hamiltonian given in Section \ref{sec_mol_hamiltonian}, together with an
extensive use of Racah's algebra and the properties outlined in Appendix
\ref{app_prop_rotation} (see \citarNP{brown_carrington03} for details). If any of the terms is not diagonal
in one of the quantum numbers of Hund's case (a) basis functions, we indicate it by using primed and non-primed quantities
for the bra and the ket of the matrix element, respectively. For instance, the rotational Hamiltonian given by 
Eq. (\ref{eq_5_30}) is diagonal in $\Lambda$, $S$, $J$ and $M$, but not in $\Sigma$ and $\Omega$.

\subsection{Spin-orbit coupling Hamiltonian}
\begin{equation}\label{eq_5_29}
\langle \Lambda S \Sigma \Omega J M | H_\mathrm{SO} | \Lambda S \Sigma \Omega J
M \rangle =
A \Lambda \Sigma + A_D \Lambda \Sigma \left[ J(J+1) - \Omega^2 + S(S+1) -
\Sigma^2 \right],
\end{equation}
\subsection{Spin-spin coupling Hamiltonian}
\begin{equation}\label{eq_5_29b}
\langle \Lambda S \Sigma \Omega J M | H_\mathrm{SS} | \Lambda S \Sigma \Omega J
M \rangle =
\frac{2}{3} \lambda \left[ 3 \Sigma^2 - S(S+1) \right],
\end{equation}
\subsection{Rotational Hamiltonian}
\begin{eqnarray}\label{eq_5_30}
\langle \Lambda S \Sigma' \Omega' J M | & H_\mathrm{rot} | \Lambda S \Sigma
\Omega J M \rangle =
B \delta_{\Sigma' \Sigma} \delta_{\Omega' \Omega} \left[ J(J+1) - \Omega^2 +
S(S+1) - \Sigma^2 
\right] \nonumber \\
&- 2B \sum_{q=\pm 1} (-1)^{J-\Omega'+S-\Sigma'}
\threej{J}{1}{J}{-\Omega'}{q}{\Omega} \threej{S}{1}{S}{-\Sigma'}{q}{\Sigma}
\nonumber  \\
&\times \sqrt{J(J+1)(2J+1)S(S+1)(2S+1)}.
\end{eqnarray}
\subsection{Centrifugal distortion Hamiltonian of order 4}
\begin{eqnarray}\label{eq_5_31}
\langle & \Lambda S \Sigma' \Omega' J M | H_\mathrm{cd}^{(4)} | \Lambda S \Sigma
\Omega J M \rangle =
 -D \delta_{\Sigma' \Sigma} \delta_{\Omega' \Omega} \left[ J(J+1) - \Omega^2 +
S(S+1) - \Sigma^2 \right]^2 \nonumber \\
& + 4D \delta_{\Sigma' \Sigma} \delta_{\Omega' \Omega} \sum_{q=\pm 1}
\sum_{\Omega'' \Sigma''}
\threej{J}{1}{J}{-\Omega}{q}{\Omega''}^2
\threej{S}{1}{S}{-\Sigma}{q}{\Sigma''}^2  \nonumber \\
&\times J(J+1)(2J+1)S(S+1)(2S+1) - 2D \sum_{q=\pm 1} (-1)^{J-\Omega'+S-\Sigma'} 
\nonumber \\
&\times \threej{J}{1}{J}{-\Omega'}{q}{\Omega}
\threej{S}{1}{S}{-\Sigma'}{q}{\Sigma} 
\sqrt{J(J+1)(2J+1)S(S+1)(2S+1)} \nonumber \\
&\times \left[ 2J(J+1) - \Omega^2 - (\Omega')^2 + 2S(S+1) - \Sigma^2 -
(\Sigma')^2 \right].
\end{eqnarray}
\subsection{Centrifugal distortion Hamiltonian of order 6}
\begin{eqnarray}\label{eq_5_32}
\langle & \Lambda S \Sigma' \Omega' J M | H_\mathrm{cd}^{(6)} | \Lambda S \Sigma
\Omega J M \rangle =
H \delta_{\Sigma' \Sigma} \delta_{\Omega' \Omega} \left[ J(J+1) - \Omega^2 +
S(S+1) - \Sigma^2 \right]^3 \nonumber \\
& + 4H \delta_{\Sigma' \Sigma} \delta_{\Omega' \Omega} \sum_{q=\pm 1}
\sum_{\Omega'' \Sigma''}
\threej{J}{1}{J}{-\Omega}{q}{\Omega''}^2
\threej{S}{1}{S}{-\Sigma}{q}{\Sigma''}^2  \nonumber \\
& \times J(J+1)(2J+1)S(S+1)(2S+1) \left[ 3J(J+1) - 2\Omega^2 - (\Omega'')^2 +
3S(S+1) - 2\Sigma^2 - (\Sigma'')^2 \right] \nonumber \\
& - 2H \sum_{q=\pm 1} (-1)^{J-\Omega'+S-\Sigma'}
\threej{J}{1}{J}{-\Omega'}{q}{\Omega} \threej{S}{1}{S}{-\Sigma'}{q}{\Sigma} \nonumber \\
&\times \sqrt{J(J+1)(2J+1)S(S+1)(2S+1)} \Bigg[ \left[ J(J+1) - \Omega^2 + S(S+1)
- \Sigma^2 \right]^2  \nonumber \\
&+ \left[ J(J+1)-(\Omega')^2 + S(S+1) - (\Sigma')^2 \right]^2 \nonumber  \\
&+\left[ J(J+1)-\Omega^2 + S(S+1) - \Sigma^2 \right] \left[ J(J+1)-(\Omega')^2 +
S(S+1) - (\Sigma')^2 \right]  \nonumber \\
&+ 4 \threej{J}{1}{J}{-\Omega'}{q}{\Omega}^2
\threej{S}{1}{S}{-\Sigma'}{q}{\Sigma}^2 J(J+1)(2J+1)S(S+1)(2S+1) \Bigg]
\end{eqnarray}
\subsection{Spin-rotation interaction Hamiltonian}
\begin{eqnarray}\label{eq_5_33}
\langle & \Lambda S \Sigma' \Omega' J M | H_\mathrm{sr} | \Lambda S \Sigma
\Omega J M \rangle =
\gamma \delta_{\Sigma' \Sigma} \delta_{\Omega' \Omega} \left[ \Omega \Sigma -
S(S+1) \right] + \gamma
\sum_{q=\pm 1} (-1)^{J-\Omega'+S-\Sigma'} \nonumber \\
&\times \threej{J}{1}{J}{-\Omega'}{q}{\Omega}
\threej{S}{1}{S}{-\Sigma'}{q}{\Sigma} \sqrt{J(J+1)(2J+1)S(S+1)(2S+1)}
\end{eqnarray}
\subsection{Zeeman Hamiltonian}
\begin{eqnarray}\label{eq_5_34}
& \langle \Lambda S \Sigma' J' M_J \Omega' | H_{\mathrm{Z}} | \Lambda S \Sigma J
M_J \Omega \rangle =
\mu_B B_0  \sum_{q=0,\pm 1} (-1)^{2J'-M-\Omega'} \left[ (2J'+1)(2J+1)
\right]^{1/2} \nonumber \\
& \times \threej{J'}{1}{J}{-M}{0}{M} \threej{J'}{1}{J}{-\Omega'}{q}{\Omega}
\Bigg[ g_L \Lambda \delta_{\Sigma \Sigma'} +
(g_S+g_r) (-1)^{S-\Sigma'} \nonumber \\
& \times \threej{S}{1}{S}{-\Sigma'}{q}{\Sigma} \left[ S(S+1)(2S+1) \right]^{1/2}
\Bigg]
-g_r \mu_B B_0 M \delta_{J J'} \delta_{\Sigma \Sigma'} \delta_{\Omega \Omega'}
\end{eqnarray}


\end{document}